\documentclass[preprint,preprintnumbers,superscriptaddress,amsmath]{revtex4}

\usepackage[dvips]{graphics}
\usepackage{epsfig}
\usepackage{dcolumn}
\usepackage{bm}
\usepackage{psfrag}
\usepackage[psamsfonts]{amssymb}

\newcommand{\be}{\begin{equation}}
\newcommand{\ee}{\end{equation}}
\newcommand{\ba}{\begin{eqnarray}}
\newcommand{\ea}{\end{eqnarray}}
\newcommand{\bs}{\boldsymbol}

\begin{document}

\preprint{APS preprint}

\title{Power law distribution of seismic rates: theory and data}

\author{A. Saichev}
\affiliation{Mathematical Department,
Nizhny Novgorod State University, Gagarin prosp. 23,
Nizhny Novgorod, 603950, Russia}
\affiliation{Institute of Geophysics and Planetary Physics,
University of California, Los Angeles, CA 90095}

\author{D. Sornette}
\affiliation{Institute of Geophysics and Planetary Physics
and Department of Earth and Space Sciences,
University of California, Los Angeles, CA 90095}
\affiliation{Laboratoire de Physique de la Mati\`ere Condens\'ee,
CNRS UMR 6622 and Universit\'e de Nice-Sophia Antipolis, 06108
Nice Cedex 2, France}
\email{sornette@moho.ess.ucla.edu}

\date{\today}

\begin{abstract}
We report an empirical determination of the
probability density functions $P_{\text{data}}(r)$
of the number $r$ of earthquakes in finite space-time windows
for the California catalog, over fixed spatial boxes $5 \times 5$ km$^2$
and time intervals $dt =1,~10,~100$ and $1000$ days. We find
a stable power law tail $P_{\text{data}}(r) \sim 1/r^{1+\mu}$ with exponent
$\mu \approx 1.6$ for all time intervals. These observations are
explained by a simple stochastic branching process previously studied by
many authors, the ETAS (epidemic-type aftershock sequence) model which
assumes that each earthquake can trigger other earthquakes
(``aftershocks''). An aftershock sequence results in this model from the
cascade of aftershocks of each past earthquake. We develop the full
theory in terms of generating functions for describing the space-time
organization of earthquake sequences and develop several approximations
to solve the equations. 
The calibration of the theory to the empirical observations
shows that it is essential to augment the ETAS model by taking account
of the pre-existing frozen heterogeneity of spontaneous earthquake
sources. This seems natural in view of the complex multi-scale nature of
fault networks, on which earthquakes nucleate.
Our extended theory is able to account for the empirical
observation satisfactorily. In particular, the adjustable parameters are
determined by fitting the largest time window $dt=1000$ days and are then used
as frozen in the formulas for other time scales, with very good 
agreement with the empirical data.
\end{abstract}


\maketitle

\section{Introduction}

Many papers purport to characterize the space-time
organization of seismicity in different regions of the
world. Recent claims of universal laws for the distribution
of waiting times and seismic rates between 
earthquakes have derived from the analyses of 
space-time windows \cite{BaketalOmo,Corral}.
The flurry of interest from physicists comes from
their fascination with the self-similar properties
exhibited by seismicity (Gutenberg-Richter power law of
earthquake seismic moments, Omori decay law of
aftershock rates, fractal and multifractal space-time
organization of earthquakes and faults) together with
the development of novel concepts and techniques that
may provide new insights
\cite{Mega,network,paczuski,paczuski2,paczuski3}. 

The interest is no less vivid among seismologists and
geophysicists in characterizing the space-time
properties of seismicity, because it allows them to
understand the dynamics of plate motion (at large
scales), to constrain the interaction between faults, as
well as to develop better hazard assessment. Recently,
an additional incentive is provided by the development
of forecasting models of seismicity, for instance within
the RELM (Regional Earthquake Likelihood Models:
www.relm.org) project in Southern California. In the
RELM project, a forecast is expressed as a vector of
earthquake rates specified for each multi-dimensional
bin \cite{Relm}, where a bin is defined by an interval
of location, time, magnitude and focal mechanism and the
resolution of a model corresponds to the bin sizes.
Then, expectations and likelihoods can be estimated and
used for the comparison between different forecasting
methods.

A fundamental issue is testing models' prediction is to
take into account so-called aftershock clustering. In
one way or another, most if not all models use some form
of declustering approach to remove the effect of
aftershocks which otherwise dominate and obscure the
desired information about the model's performance
\cite{Relm}. Then, with such declustered catalogs, the
likelihood of forecasts are estimated using Poissonian
probabilities. But, if the catalog is only partially
declustered (which it will most probably be as there are
no agreed upon fully efficient method of declustering),
then our contribution in this paper is to show that the
distribution of event numbers should present a tail much
more heavy than predicted by the Poissonian statistics
and to propose a theoretical explanation for it.
Pisarenko and Golubeva \cite{PisGol} introduced a model
to decluster catalogs by so-to-say Poisson ``with random
parameter,'' which resulted in a law with slowly
decreasing probabilities (actually a stable L\'evy law)
for the distribution of rates. 

We improve on
preceding results on several points. Our first contribution
is to show that the heavy tail nature of the
distribution of seismic rate is intrinsic to a class of
generic models of triggered seismicity. Specifically,
our theory is based on a simple model of earthquake
triggering, in which future seismicity is a conditional
Poisson process, with average rates (or Poisson
intensity) conditioned on past seismicity. We show that
the exponential Poisson rate is renormalized into a
power law tail by the mechanism involving a cascade of
earthquake triggering. Our theory thus provides a
prediction for the distribution of seismic rates in
space-time bins in the form of a power law tail
distribution. Our second contribution is to show that
our prediction is verified by empirical seismic rates in
Southern California over more than two decades. In
addition, our theory accounts well for the evolution of
the distribution of seismic rates as a function of the
time window size from 1 day to 1000 days.

This implies that spontaneous fluctuations of the number of triggered
earthquakes
in space-time bins
may be simply due to the cascades of triggering  processes, which lead
to dramatic departures from the Poisson model used as one of the
building block of
standard testing procedures. Accounting for the intrinsic heavy tail
nature of the distribution of seismic rate may explain, we believe, many
of the contradictions and rejections of models assessed on the basis of Poisson
statistics of so-called declustered catalogs. This also suggests
the need for fundamentally different earthquake prediction models and
testing methods. Our results also offer a simple alternative
explanation to so-called universal laws \cite{BaketalOmo,Corral} in terms of
cascades of triggered earthquakes: our proposed
framework explains the observed power law distributions
of seismic rates from the fundamentals of seismicity
characterized by a few exponents.

\section{The Epidemic-Type Aftershock Sequence (ETAS) branching model
of earthquakes with long memory \label{laws}}

We study the general branching process, called the Epidemic-Type
Aftershock Sequence (ETAS) model of triggered seismicity, introduced by
Ogata in the present form \cite{Ogata} and by Kagan and Knopoff in a
slightly different form \cite{KK81} and whose main statistical
properties are reviewed in \cite{HS02}. For completeness and in order to fix
notations, we recall its
definition and ingredients used in our analysis that follows.
In this model, all earthquakes are treated on the same footing
and there is no distinction between foreshocks,
mainshocks and aftershocks, other than
from retrospective human-made classification.
The advantage of the ETAS
model is its conceptual simplicity based on three independent
well-found empirical laws and its power of explanation of other
empirical observations (see for instance
\cite{Forexp} and references therein).

The ETAS model belongs to a general class of branching processes
\cite{Athreya,Sankaranarayanan}, and has in addition the property that
the variance of the number of earthquake progenies triggered in
direct lineage from
a given mother earthquake is mathematically infinite. Moreover, a
long-time (power
law) memory of the impact of a mother on her
first-generation daughters describes the empirical Omori law for aftershocks.
These two ingredients together with the mechanism of cascades of branching
have been shown to give rise to subdiffusion
\cite{etasdif,HOS03} and to non mean-field behavior in the
distribution of the total number of aftershocks per
mainshock, in the
distribution of the total number of generations before
extinctions \cite{Saichevetal04} and in the
distribution of the total duration of an
aftershock sequence before extinction \cite{SaichSorl04}.

In the ETAS model, each earthquake is a potential
progenitor or mother, characterized by its conditional
average number \be N_m \equiv \kappa \mu(m)
\label{avenun} \ee of children (triggered events or
aftershocks of first generation), where \be \mu(m) =
10^{\alpha (m-m_0)}~, \label{mudef} \ee is a mark
associated with an earthquake of magnitude $m \geqslant
m_0$ (in the language of ``marked point processes''),
$\kappa$ is a constant factor and $m_0$ is the minimum
magnitude of earthquakes capable of triggering other
earthquakes. The meaning of the term ``conditional
average'' for $N_m$ is the following: for a given
earthquake of magnitude $m$ and therefore of mark
$\mu(m)$, the number $r$ of its daughters of first
generation are drawn at random according to the
Poissonian statistics \be p_\mu(r)=
\frac{N_m^r}{r!}\,e^{-N_m} =
\frac{(\kappa\mu)^r}{r!}\,e^{-\kappa\mu}\,. \label{3aa}
\ee $N_m$ is the expectation of the number of daughters
of first generation, conditioned on a fixed magnitude
$m$ and mark $\mu(m)$. The expression (\ref{mudef}) for
$\mu(m)$ is chosen in such a way that it reproduces the
empirical dependence of the average number of
aftershocks triggered directly by an earthquake of
magnitude $m$ (see \cite{alpha} and references therein).
Expression (\ref{avenun}) with (\ref{mudef}) gives the
so-called productivity law of a given mother as a
function of its magnitude. The challenge of our present
analysis is to understand how the exponential
distribution (\ref{3aa}) is changed by taking into
account all earthquake triggering paths simultaneously
and at all possible generations.

The ETAS model is complemented by the Gutenberg-Richter
(GR) density distribution of earthquake magnitudes \be
p(m) = b ~\ln (10)~ 10^{-b (m-m_0)}~,~~~~m \geqslant
m_0~, \label{GR} \ee such that $\int_m^{\infty} p(x) dx$
gives the probability that an earthquake has a magnitude
equal to or larger than $m$. This magnitude distribution
$p(m)$ is assumed to be independent of the magnitude of
the triggering earthquake, i.e., a large earthquake can
be triggered by a smaller one \cite{alpha,Forexp}.

Combining (\ref{GR}) and (\ref{mudef}) shows that the
earthquake marks $\mu$ and therefore the conditional average
number $N_m$ of daughters of first generation are
distributed according to the normalized power law
\be
p_{\mu}(\mu) = {\gamma \over \mu^{1+\gamma}}~,
~~~1 \leq \mu < +\infty, ~~~~~\gamma = b/\alpha~.
\label{aera}
\ee
For earthquakes, $b \approx 1$ and $0.5 < \alpha < 1$
giving $1 < \gamma <2$ (see \cite{SorWer} for a review
of values quoted in the literature and their implications).
This range $1 < \gamma <2$ implies that the
mathematical expectation of $\mu$ and therefore of $N_m$
(performed over all possible magnitudes) is finite but its variance
is infinite (the marginal case $\alpha = 1$
leading to $\gamma =1$ requires the existence of an upper magnitude
cut-off \cite{SorWer}).

For a fixed $\gamma$, the coefficient $\kappa$ then controls the value of the
average number $n$ (or branching ratio) of children of first
generation per mother:
\be
n = \langle N_m \rangle = \kappa \langle \mu \rangle = \kappa {\gamma
\over \gamma -1}~,
\label{mgmlele}
\ee
where the average $\langle N_m \rangle$ is taken over all mothers' magnitudes
drawn from the GR law. Recall that the values $n<1$, $n=1$ and $n>1$
correspond respectively to the sub-critical, critical and
super-critical regimes.

The next ingredient of the ETAS model consists in the
specification of the space-time rate function $N_m
~\Phi(\bs{r}-\bs{r}_i,t-t_i)$ giving the average rate of
first generation daughters at time $t$ and position
$\bs{r}$ created by a mother of magnitude $m\geqslant
m_0$ occurring at time $t_i$ and position $\bs{r}_i$:
\be \Phi(\bs{x},t)=\Phi(t)\, \phi(\bs{x})\,.
\label{jfaphapa} \ee The time propagator $\Phi(t)$ has
the Omori law form
\begin{equation}
\label{Omori}
\Phi(t)= \frac{\theta c^\theta}{(c+t)^{1+\theta}}~H(t)
\end{equation}
where $H(t)$ is the Heaviside function, $0<\theta<1$, $c$
is a regularizing time scale that
ensures that the seismicity rate remains finite close to
the mainshock. The
time decay rate (\ref{Omori}) is called the ``direct
Omori law''  \cite{SS99,HS02}. Due to the process of
cascades of triggering by which a mother triggers
daughters which then trigger their own daughters and so
on, the direct Omori law (\ref{Omori}) is renormalized
into a ``dressed'' or ``renormalized'' Omori law
\cite{SS99,HS02}, which is the one observed empirically. The analysis
below will retrieve and extend this result.

The space propagator is given by
\begin{equation}
\label{9}
\phi(\bs{x})= \frac{\eta ~d^{\eta}}{2 \pi (x^2+d^2)^{(\eta+2)/2}}\,.
\end{equation}
For our comparison with the empirical data,
we shall consider the epicenter position of earthquakes, that is, the
2D-projection on the earth surface of the real 3D-distribution of
earthquake hypocenters. Numerical implementations of the theory developed below
will thus be  done in 2D but it is easy to generalize to 3D if/when the
empirical data will be of sufficient quality to warrant it.

In the following, we will make use of the Laplace
transform of the Omori law \be
\hat{\Phi}(u)=\int_0^\infty \Phi(t)\, e^{-ut}\, dt=
\theta\,(cu)^\theta\,e^{cu}\, \Gamma(-\theta,cu) \ee and
of its asymptotic behavior
\begin{equation}
\label{10}
\hat{\Phi}^{-1}(u)\sim 1+ \Gamma(1-\theta)(cu)^\theta\,,
\qquad cu\ll 1\,.
\end{equation}
The Fourier transform of the space propagator (\ref{9}) will also be useful:
\be
\tilde{\phi}(\bs{q})=
\iint\limits_{-\infty}^{\quad\infty} \phi(\bs{x})
e^{i(\bs{q}\cdot\bs{x})}\,d\bs{x} = 2
\left(\frac{dq}{2}\right)^{\eta/2} \frac{K_{\eta/2}(dq)}
{\Gamma\left({\eta/2}\right)}
\ee
In particular
\be
\tilde{\phi}(\bs{q})= e^{-dq} \quad (\eta=1)\,, \qquad
\tilde{\phi}(\bs{q})= e^{-dq}(1+dq) \quad (\eta=3)\,.
\ee

The last ingredient of the ETAS model is to assume that plate tectonic
motion induces spontaneous mother earthquakes, which are not triggered by
previous earthquakes, according to a Poissonian point process, such
that the average number of spontaneous mother earthquakes per unit time
and per unit surface is $\varrho$. In the
ETAS branching model, each such spontaneous mother earthquake then triggers
independently its own space-time aftershocks branching process.

It is a well-established facts that, at large scale, earthquakes are
preferentially clustered near the plate boundaries while, at smaller scales,
earthquakes are found mostly along faults and close to nodes between
several faults \cite{russ}. It is natural to extend the ETAS model to allow for
the heterogeneity of the spontaneous earthquake sources $\varrho$ reflecting
the influence of pre-existing fault structures, some
rheological heterogeneity and complex spatial stress distributions.
We get some guidelines from the distribution of the stress field
in heterogeneous media and due to earthquakes \cite{Kaganstress}
which should be close to a Cauchy distribution
or probably more generally to a power law \cite{Zolo1,Zolo2} (see also
Chap.~17 of \cite{Sorbook1}).

The simplest prescription is thus to assume that
$\varrho$ is itself random and distributed according to
\be \frac{1}{\langle\varrho\rangle}\,
f\left(\frac{\varrho}{\langle\varrho\rangle}\right)~,
\label{disrholl} \ee where $\langle\varrho\rangle$ is
then statistical average of the random space-time
Poissonian source intensity $\varrho$. In the numerical
applications below, we shall use the form \be
f_\delta(x)= \frac{\delta+1}{\delta} \left(1+
\frac{x}{\delta}\right)^{-2-\delta}\qquad (\delta>0)\,,
\label{disrholl2} \ee The value $\delta=0$ gives the
same tail as the Cauchy distribution advocated in
\cite{Kaganstress} for the stress field. We have
considered other functions, such as half-Gaussian,
exponential, half-Cauchy but none of them give
satisfactory fits to the data (see below). The
parametrization (\ref{disrholl2}) with $\delta>0$ allows
us to have only a single scale $\langle\varrho\rangle$
controlling the typical fluctuation of the random
sources. We have found that only slightly positive values of
$\delta$ (corresponding to tails a little fatter than the Cauchy
law) gives reasonable fits to the data (see below). It is interesting
to observe that the data on the distribution of seismic rates 
thus seems to constrain significantly the fractal 
distribution of seismic sources.

\section{Generating Probability Function (GPF) of earthquakes
branching process}

In this section, we describe the statistical
properties of earthquake branching processes using the technology
of generating functions.

First, let us recall the GPF of the total number $R_1$ of the first-generation
aftershocks of a mother event of magnitude $m$ can be easily obtained as
\begin{equation}\label{11}
e^{\mu\kappa(z-1)}\,,
\end{equation}
using the fact that the rate of first-generation aftershocks is
Poissonian according
to (\ref{3aa}). In this expression (\ref{11}), $\kappa$ and $\mu$ are
given by their
definition in (\ref{avenun}) and (\ref{mudef}).

Averaging (\ref{11}) over the random parameter $\mu$ gives the
GPF of the number $R_1$ of first generation aftershocks
triggered by a mother aftershock of arbitrary magnitude
\begin{equation}\label{12}
G(z)=\gamma\kappa^\gamma(1-z)^\gamma\,
\Gamma(-\gamma,\kappa(1-z))\,.
\end{equation}
Note that $\left<R_1\right>$ is nothing but the branching ratio $n$
defined above in equation (\ref{mgmlele}).
Knowing the expression (\ref{12}) for the GPF $G(z)$, one finds
that the corresponding probabilities of the random
numbers $R_1$ are equal to
\begin{equation}
\label{13}
P_1(r)=\text{Pr\,}\{R_1=r\}= \gamma\frac{k^\gamma}{r!}
\Gamma(r-\gamma,\kappa)~,
\end{equation}
which have the following asymptotics
\begin{equation}
\label{14}
P_1(r)\simeq \frac{\gamma\kappa^\gamma}{r^{\gamma+1}}=
n^\gamma\, \gamma^{1-\gamma}\, (\gamma-1)^\gamma\,
r^{-\gamma-1}\qquad (r\gg 1)\,.
\end{equation}
Expression (\ref{14}) implies that, for $1<\gamma<2$, the variance of
the random
number $R_1$ is infinite. For $\gamma > 2$, the variance is finite
and is equal to
\be
\sigma^2_1= \frac{n^2}{\gamma(\gamma-2)}+ n\,.
\ee
Figure~1 shows the probabilities
(\ref{13}) and their power law asymptotics for $n=1$,
$\gamma=1.25$ and $\gamma=3$.

Let us now consider the set of independent space-time aftershock
branching processes,
triggered by spontaneously arising mother earthquakes. Due to the
independence between each sequence triggered by each spontaneous
event, it is easy
to show that the GPF of the number of
events (including mother earthquakes and all their aftershocks of all
generations),
falling into the space-time window $\{[t,t+\tau]\times
\mathcal{S}\}$ is equal to
\begin{equation}\label{1}
\Theta_{\text{sp}}(z,\tau,\mathcal{S})=e^{-\varrho\,
L(z,\tau,\mathcal{S})}
\end{equation}
where
\begin{equation}\label{2}
\begin{array}{c}
\displaystyle L(z,\tau,\mathcal{S})=\int_0^\infty dt
\iint\limits_{-\infty}^{\quad\infty} d\bs{x}
[1-\Theta(z,t,\tau,\mathcal{S};\bs{x})]+\\[2mm]
\displaystyle \int_0^\tau dt
\iint\limits_{-\infty}^{\quad\infty} d\bs{x}
[1-\Theta(z,t,\mathcal{S};\bs{x})][1-I_\mathcal{S}(\bs{x})]+\\[5mm]
\displaystyle \int_0^\tau dt \iint\limits_{\mathcal{S}}
d \bs{x}[1-z\Theta(z,t,\mathcal{S};\bs{x})]\,.
\end{array}
\end{equation}
The three above summands have the following transparent geometrical meaning.
\begin{itemize}
\item The first summand describes the contribution to the GPF
$\Theta_{\text{sp}}$ from aftershocks triggered
by mother earthquakes that occurred before the time window
(i.e. at instants $t'$ such that $t'<t$) (positions $1$ in Fig.~2).
The corresponding GPF
$\Theta(z,t-t',\tau,\mathcal{S};\bs{x})$ of the
number of aftershocks triggered inside the space-time window
$\{[t,t+\tau]\times \mathcal{S}\}$ by some mother event that occurred
at time $t'$ satisfies the relation
\begin{equation}
\label{3bb}
\Theta(z,t,\tau,\mathcal{S};\bs{x})=
G[1-\Psi(z,t,\tau,\mathcal{S};\bs{x})]\,,
\end{equation}
where the auxiliary function
$\Psi(z,t,\tau,\mathcal{S};\bs{x})$, describing the
space-time dissemination of aftershocks triggering by
some mother event, is equal to
\begin{equation}
\label{4}
\begin{array}{c}
\displaystyle \Psi(z,t,\tau,\mathcal{S};\bs{x})=\\[2mm]
\Phi(\bs{x},t) \otimes [1-
\Theta(z,t,\tau,\mathcal{S};\bs{x})]+
\Phi(\bs{x},t+\tau)\otimes [1-
\Theta(z,\tau,\mathcal{S};\bs{x})]+
\\[2mm]
(1-z)\Phi(\bs{x},t+\tau)\otimes I_\mathcal{S}(\bs{x})
\Theta(z,\tau,\mathcal{S};\bs{x})\,.
\end{array}
\end{equation}
$\Phi(\bs{x}-\bs{x}',t')$, which has been defined in (\ref{jfaphapa}),
is the probability density function (pdf) of the position
$\bs{x}'$ and instant $t'$ of some first generation
aftershock, triggered by the mother event, arising at
the instant $t=0$ and at the point $\bs{x}$.
The function $I_\mathcal{S}(\bs{x})$ in (\ref{4}) is
the indicator of the space window $\mathcal{S}$ and $G(z)$ in (\ref{3bb})
is the GPF of the number $R_1$ of first generation
aftershocks, triggered by some mother earthquake. $G(z)$ given in (\ref{12})
is common to all mother earthquakes and to all aftershocks.

\item The last two terms in expression (\ref{2})
(positions $2$ and $3$ in Fig.~2) describe the
contribution of aftershocks triggered by earthquakes,
occurring inside the time window (i.e., $t'\in[t,t+\tau]$).
The second term (position $2$ in Fig.~2) corresponds to the subset
spatially outside the domain $\mathcal{S}$.
The third term (position $3$ in Fig.~2) corresponds to the subset
spatially inside the domain $\mathcal{S}$.
These last two terms in expression (\ref{2}) depend on the GPF
\be
\Theta(z,\tau,\mathcal{S};\bs{x})=
\Theta(z,t=0,\tau,\mathcal{S};\bs{x})
\ee
of the numbers of aftershocks triggered till time $\tau$ inside
the space window $\mathcal{S}$ by some mother event
arising at the instant $t=0$ and at the point $\bs{x}$.
It follows from (\ref{3bb}) and (\ref{4}) that it satisfies
the relations
\begin{equation}\label{5}
\Theta(z,\tau,\mathcal{S};\bs{x})=
G[1-\Psi(z,\tau,\mathcal{S};\bs{x})]
\end{equation}
and
\begin{equation}\label{6}
\begin{array}{c}
\Psi(z,\tau,\mathcal{S};\bs{x})=\\
  \Phi(\bs{x},\tau)\otimes [1-
\Theta(z,\tau,\mathcal{S};\bs{x})]+
(1-z)\Phi(\bs{x},\tau)\otimes I_\mathcal{S}(\bs{x})
\Theta(z,\tau,\mathcal{S};\bs{x})\,.
\end{array}
\end{equation}
\end{itemize}

In addition, we shall need the GPF
\be
\Theta(z,\mathcal{S};\bs{x})=
\Theta(z,\tau=\infty,\mathcal{S};\bs{x})
\ee
of the total numbers of aftershocks triggered by some mother
earthquake inside the area $\mathcal{S}$. As seen
from (\ref{5}) and (\ref{6}), it satisfies the relations
\begin{equation}\label{7}
\Theta(z,\mathcal{S};\bs{x})=
G[1-\Psi(z,\mathcal{S};\bs{x})]
\end{equation}
and
\begin{equation}\label{8}
\Psi(z,\mathcal{S};\bs{x})= 1- \phi(\bs{x})\otimes
\Theta(z,\mathcal{S};\bs{x})+ (1-z) \phi(\bs{x})\otimes
I_\mathcal{S}(\bs{x}) \Theta(z,\mathcal{S};\bs{x})\,.
\end{equation}

Taking into account the distribution of the source intensities $\varrho$
amounts to averaging equation (\ref{1}) over $\varrho$ weighted with
the statistics
(\ref{disrholl}). This gives
\begin{equation}
\label{15}
\Theta_{\text{sp}}(z,\tau;\mathcal{S})=\hat{f}[\langle\varrho\rangle\,
L(z,\tau,\mathcal{S})]\,,
\end{equation}
where $\hat{f}(u)$ is the Laplace transform of the pdf $f(x)$.
For the specification (\ref{disrholl2}), expression (\ref{15})
becomes
\be
\hat{f}_\delta(u)= (1+\delta) (\delta u)^{1+\delta}\,
e^{\delta u}\, \Gamma(-1-\delta,\delta u)\,.
\ee

\section{Averages and rates of aftershocks within the space-time window
$\{[t,t+\tau]\times \mathcal{S}\}$}

Before discussing the properties of the distributions of aftershocks, we
consider their simplest statistical characteristics, namely the averages
and rates of different kinds of aftershocks. This introduces the
relevant characteristic scales in the time and in the space domains,
which are found inherent to the space-time branching processes. This
also suggests the natural ``large time window approximation'' used
and tested below within the more general probabilistic treatment.

\subsection{Average of the total number of events in the
space time window $\{[t,t+\tau]\times \mathcal{S}\}$}

Let us first calculate the average the total number of events
inside the space-time window given by
\be
\langle R_{\text{sp}}(\tau,\mathcal{S})\rangle = \left.
\frac{\partial \Theta_{\text{sp}}(z,\tau,\mathcal{S})}{\partial
z}\right|_{z=1}~.
\ee
It follows from (\ref{15}) and (\ref{2}) that it is equal to
\be
\langle R_{\text{sp}}(\tau,\mathcal{S})\rangle = \langle
R_{\text{out}}(\tau,\mathcal{S})\rangle + \langle
R(\tau,\mathcal{S})\rangle + \langle \varrho\rangle  S
\tau\,,
\ee
where $\langle R_{\text{out}}(\tau,\mathcal{S})\rangle$
is the average number of aftershocks triggered by
spontaneous ``mother'' earthquake sources that occurred before time $t$
(positions $1$ in Fig.~2),
$\langle R(\tau,\mathcal{S})\rangle$
is the average of number of aftershocks triggering by spontaneous
earthquake sources
that occur within the time interval $[t,t+\tau]$ (positions $2$ and
$3$ in Fig.~2)
and $\langle\varrho\rangle S\tau$ is the
average of number of spontaneous earthquakes inside the
space-time window. Here and everywhere in the following, $S$ is the area
of the spatial domain $\mathcal{S}$ and thus $S \tau$ is the space-time volume
associated with the space-time window.

In what follows, it will be useful to introduce the rate of events
\be
N_{\text{sp}}(\tau,\mathcal{S})= \frac{d \langle
R_{\text{sp}}(\tau,\mathcal{S})\rangle}{d\tau}~,
\ee
with
\begin{equation}
\label{16}
N_{\text{sp}}(\tau,\mathcal{S})=
N_{\text{out}}(\tau,\mathcal{S})+ N(\tau,\mathcal{S})+
\langle \varrho\rangle  S\,.
\end{equation}
Using Eq. (\ref{3bb}), (\ref{4}) and Eq. (\ref{5}),
(\ref{6}), one shows that
\begin{equation}\label{17}
N(\tau,\mathcal{S}) =\frac{n}{1-n} \langle
\varrho\rangle \,S- N_{\text{out}}(\tau,\mathcal{S}) \,.
\end{equation}
where $N_{\text{out}}(\tau,\mathcal{S})$ satisfies the
equation
\begin{equation}\label{18}
N_{\text{out}}(\tau,\mathcal{S}) - n\,
N_{\text{out}}(\tau,\mathcal{S}) \otimes \Phi(\tau)=
\frac{n\,\langle \varrho\rangle S}{1-n} a(\tau)\,,
\end{equation}
and $n$ is the branching ratio defined in (\ref{mgmlele}).
Here and below, the following notation is used
\be
a(\tau)=\int_\tau^\infty \Phi(t)\, dt=
\left(\frac{c}{\tau+c}\right)^\theta\,.
\label{kkdksd}
\ee
Substituting (\ref{17}) into (\ref{16}) gives the obvious
equality
\begin{equation}
\label{19}
N_{\text{sp}}(\tau,\mathcal{S}) = \frac{\langle
\varrho\rangle S}{1-n}~,
\end{equation}
which implies that, due to the cascade of earthquake triggering processes,
the average of the total number $\langle
R_{\text{sp}}(\tau,\mathcal{S})\rangle$
of events is amplified by the factor $1/(1-n)$ compared with the
average number $\langle \varrho\rangle\tau S$ of
earthquake sources. This factor $1/(1-n)$ has a simple intuitive
meaning \cite{HS03n}:
one event gives on average $n$ daughters in direct lineage; each of
these first-generation daughters give $n$ grand-daughters,
the average number of grand-daughters is thus $n^2$, and the
reasoning continues
over all generations. Summing over all generations,
the total number of events triggered by a given source plus the source itself
is $1+ n + n^2 + n^3 + ...$, which sums to $1/(1-n)$.

\subsection{Impact of mother earthquake sources occurring before the
time window}

Here, we use previous and other related relations in the goal of
estimating the contribution of the different
terms in the r.h.s. of Eq.~(\ref{2}) to the GPF
$\Theta_{\text{sp}}(z,\tau,\mathcal{S})$. Our goal is to prepare
and check for approximations that will be used below.

For instance, we shall assume that the contribution of the first term
in (\ref{2}),
which is responsible for aftershocks triggered by earthquakes occurring
before time $t$ (i.e., outside the time window $[t,t+\tau]$), is negligible if
the corresponding relative events rate obeys the following condition
\be
\mathcal{N}_{\text{out}}(\tau)=
\frac{N_{\text{out}}(\tau,\mathcal{S})}{N_{\text{sp}}(\tau,\mathcal{S})}
\ll 1\,.
\ee
To check when this condition holds, notice that, due
to (\ref{18}), (\ref{19}),
$\mathcal{N}_{\text{out}}(\tau)$ is solution of
\be
\mathcal{N}_{\text{out}}(\tau) - n\,
\mathcal{N}_{\text{out}}(\tau)\otimes \Phi(\tau)= n a(\tau)\,.
\ee
Applying the Laplace transform to both sides of this equation gives
\be
\hat{\mathcal{N}}_{\text{out}}(u)= n
\frac{1-\hat{\Phi}(u)}{u[1-n \hat{\Phi}(u)]}\,.
\ee
Using the asymptotic formula (\ref{10}), we obtain
\begin{equation}
\label{20}
\hat{\mathcal{N}}_{\text{out}}(u)\simeq \frac{n\, (c_1
u)^{\theta-1}}{1+ (c_1 u)^\theta}\,,
\end{equation}
where
\be
c_1=
\left(\frac{\Gamma(1-\theta)}{1-n}\right)^{1/\theta}\, c
\label{ngjmss,s,}
\ee
is a characteristic time-scale of aftershock
branching processes separating a $1/t^{1-\theta}$ law at $t < c_1$ from
a $1/t^{1+\theta}$ law at $t>c_1$ for the decay with time of the average
aftershock rate triggered by a single mother earthquake \cite{SS99,HS02}.
For instance, if $c=2$ min,
$n=0.9$, $\theta=1/2$ then $c_1\simeq 628$ min $\simeq 10.5$ hours.

Taking the inverse Laplace transform of Eq.~(\ref{20}) gives
\begin{equation}
\label{21}
\mathcal{N}_{\text{out}}(\tau)= n\,
E_\theta\left[-\left(\frac{\tau}{c_1}\right)^\theta\right]
\end{equation}
where $E_\theta(x)$ is the Mittag-Leffler function defined by
\be
E_\theta(-x)= \frac{x}{\pi} \sin\pi\theta \int_0^\infty
\frac{y^{\theta-1} e^{-y}\, dy}{y^{2\theta}+ x^2+ 2 x
y^\theta \cos\pi\theta} \quad (x>0, \ 0<\theta<1)\,.
\ee
The following asymptotic property holds:
\begin{equation}
\label{22}
E_\theta(-x) \sim \frac{1}{x ~\Gamma(1-\theta)}\qquad
(x\to\infty)\,
\end{equation}
In addition,
\be
E_{1/2}(-x)= e^{x^2}\, \text{erfc\,} x\,.
\ee

Figure~3 plots the exact rate
$\mathcal{N}_{\text{out}}(\tau)$, its fractional
approximation (\ref{21}) and corresponding asymptotics derived from (\ref{22}):
\begin{equation}
\label{23}
\mathcal{N}_{\text{out}}(\tau)\simeq
\frac{n}{\Gamma(1-\theta)}\,
\left(\frac{c_1}{\tau}\right)^{\theta}~,
\end{equation}
for $n=0.9$ and $\theta=1/2$. One can observe that the asymptotic result
(\ref{23}) is rather precise even if $\tau$ is close to
$c_1$. Eq.~(\ref{23}) means that, if
\begin{equation}
\label{24}
\tau \gg c_1~,
\end{equation}
then one can neglect the contribution of aftershocks
triggered by the spontaneous earthquake sources occurring before the
time window $[t,t+\tau]$.
The remark will be used in our following investigation and
the condition (\ref{24}) will be refered to as the
``large time window approximation.''

\subsection{Impact of mother earthquake sources occurring inside
the space-time window $\{[t,t+\tau]\times \mathcal{S}\}$}

Let calculate the contribution to the rate (\ref{16}) of
events corresponding to the last term of (\ref{2}) (position $3$ in Fig.~2).
This term describes all the aftershocks triggered by
mother earthquakes occurring within the space-time
window  $\{[t,t+\tau]\times \mathcal{S}\}$. It is easy to show that the
corresponding seismic rate, which can be compared with the contribution
(\ref{19}), is equal to
\begin{equation}
\label{25}
\mathcal{N}_{\text{in}}(\tau;\mathcal{S})= 1-n+
\frac{1-n}{S} \iint\limits_{\mathcal{S}}\langle
R(\tau,\mathcal{S};\bs{x})\rangle d\bs{x}\,,
\end{equation}
where
\be
\langle R(\tau,\mathcal{S};\bs{x})\rangle= \left.
\frac{\partial \Theta(z,\tau,\mathcal{S})}{\partial
z}\right|_{z=1}
\ee
is the average number of aftershocks triggered within
the space window $\mathcal{S}$ till instant $\tau$ by
some mother earthquake occurring at position $\bs{x}$
and at the time $t=0$. Using relations
(\ref{5}) -- (\ref{8}), one can show that
\begin{equation}
\label{26}
\langle R(\tau,\mathcal{S};\bs{x})\rangle= \langle
R(\mathcal{S};\bs{x})\rangle- \langle
R_+(\tau,\mathcal{S};\bs{x})\rangle\,,
\end{equation}
where $\langle R(\mathcal{S};\bs{x})\rangle$ is the total
number of aftershocks falling inside the space domain
$\mathcal{S}$ which are triggered by a
earthquake source occurring at position $\bs{x}$
and at the time $t=0$.
$\langle R_+(\tau,\mathcal{S};\bs{x})\rangle$ is the corresponding number of
aftershocks falling with the space domain $\mathcal{S}$
after the instant $\tau$.

It is easy to show that the Laplace (with respect to $\tau$)
and the Fourier (with respect to $\bs{x}$) transform of
$\langle R_+(\tau,\mathcal{S};\bs{x})\rangle$ is equal
to
\be
\hat{\langle R\rangle}_+(u,\mathcal{S};\bs{q})= n\,
\tilde{\phi}_\mathcal{S}(\bs{q}) \frac{1}{u} \left(
\frac{1}{1-n\tilde{\phi}(\bs{q})}-
\frac{\tilde{\Phi}(u)}{1-n\tilde{\phi}(\bs{q})
\tilde{\Phi}(u)}\right)\,,
\ee
Using the asymptotics (\ref{10}), we can rewrite this last relation in
the form
\begin{equation}
\label{27}
\hat{\langle R\rangle}_+(u,\mathcal{S};\bs{q})\simeq
\langle\tilde{R}\rangle(\mathcal{S};\bs{q})\,
\frac{\Gamma(1-\theta) c^\theta\,
u^{\theta-1}}{\Gamma(1-\theta)(cu)^\theta+1-
n\tilde{\phi}(\bs{q})}\,,
\end{equation}
where $\langle \tilde{R}\rangle(\mathcal{S};\bs{q})$ is
the Fourier transform of the average
$\langle R(\mathcal{S};\bs{x})\rangle$ of the total number of
aftershocks mentioned above. Using relations (\ref{7}),
(\ref{8}), we obtain
\be
\langle \tilde{R}\rangle(\mathcal{S};\bs{q})= \frac{n\,
\tilde{\phi}(\bs{q})}{1-n\tilde{\phi}(\bs{q})}\,
\tilde{I}_\mathcal{S}(\bs{q})\,,
\ee
where $\tilde{I}_\mathcal{S}(\bs{q})$ is the
Fourier transform  of
the indicator function of the space window $\mathcal{S}$. In what follows,
we assume that $\mathcal{S}$ is the circular domain of
radius $\ell$ centered at the origin of the plane
$\bs{x}$. Then
\be
\tilde{I}_\mathcal{S}(\bs{q})= 2\pi\, \frac{\ell}{q}\,
J_1(\ell q)\,.
\label{mghkhlr}
\ee

Eq. (\ref{27}) implies that $\langle
R_+(\tau,\mathcal{S};\bs{x})\rangle$ is given by
\be
\langle R_+(\tau,\mathcal{S};\bs{x})\rangle= \langle
R(\mathcal{S};\bs{x})\rangle\otimes
\mathcal{H}(\tau;\bs{x})\,,
\ee
where the Fourier transform of the function
$\mathcal{H}(\tau;\bs{x})$ is equal to
\be
\tilde{\mathcal{H}}(\tau;\bs{q})= E_\theta\left(-
\frac{1-n \tilde{\phi}(\bs{q})}{1-n}
\left(\frac{\tau}{c_1}\right)^\theta\right)\,.
\ee
Thus, the Fourier transform (with respect to $\bs{x}$) of the
sought average given by (\ref{26}) is
\begin{equation}
\label{28}
\langle \tilde{R}\rangle(\tau,\mathcal{S};\bs{q})\simeq
\langle
\tilde{R}\rangle(\mathcal{S};\bs{q})\left[1-E_\theta\left(-
\frac{1-n \tilde{\phi}(\bs{q})}{1-n}
\left(\frac{\tau}{c_1}\right)^\theta\right)\right]\,.
\end{equation}
Using expression (\ref{28}), we construct Figure 4 which plots $\langle
R(\tau,\mathcal{S};\bs{x})\rangle$ for different values of
$\tau/c_1$, in order to illustrate its convergence to the average
$\langle R(\mathcal{S};\bs{x})\rangle$ of the total
number of aftershocks falling inside the area $\mathcal{S}$.

As can be seen from figures 3 and 4,
it follows from (\ref{28}) and from the properties of
Mittag-Leffler functions that, if the large time window approximation
(\ref{24}) holds, one may use the approximate equality
\be
\langle R(\tau,\mathcal{S};\bs{x})\rangle\simeq \langle
R(\mathcal{S};\bs{x})\rangle~.
\ee
In this large time window approximation, the relative rate
(\ref{25}) is transformed into
\begin{equation}\label{29}
\mathcal{N}_{\text{in}}(\mathcal{S})\simeq 1-n+
\frac{n(1-n)}{2\pi^2 \ell^2} \int_0^\infty
\frac{\tilde{\phi}(q)}{1-n \phi(q)}
\tilde{I}_\mathcal{S}^2(q)\, dq\,.
\end{equation}
As the space domain $\mathcal{S}$ increases in size,
$\mathcal{N}_{\text{in}}(\mathcal{S})$ increases towards $1$.
Figure~5 plots $\mathcal{N}_{\text{in}}(\mathcal{S})$ using the space
propagator $\phi(\bs{x})$ given by (\ref{9}) for different values of the
exponent $\eta$.

\section{Large time window approximation}

The analysis of the previous section gives us the possibility
to explore the probabilistic properties of the number of
events in given space-time windows, in the regime where the
large time window approximation (\ref{24}) holds. If
the time duration $\tau$ of the
space-time window is sufficiently large, the previous section has
shown that the statistical averages and the seismic rates become independent
of $\tau$. It seems reasonable to conjecture that the
GPF $\Theta(z,\tau,\mathcal{S};\bs{x})$ of the total number
of aftershocks triggered by some earthquake source inside
the space domain $\mathcal{S}$ until time $\tau$
coincides approximately with the saturated GPF
$\Theta(z,\mathcal{S};\bs{x})$ of the total number of aftershocks
triggered by some earthquake source inside the space domain
$\mathcal{S}$. Within this approximation of large time windows,
the effect of aftershocks triggered by
earthquake sources occurring till the beginning $t$ of the time window is
negligible. Section \ref{exmmflaa} below will explore
in more details the applicability of this conjecture.

Within this large time window approximation, one may ignore the first
term in the
r.h.s. of Eq.~(\ref{2}) and replace
$\Theta(z,t,\mathcal{S};\bs{x})$ by
$\Theta(z,\mathcal{S};\bs{x})$ in the remaining terms. As a result,
Eq.~(\ref{2})
takes the following approximate form
\begin{equation}
\label{30}
L(z,\tau,\mathcal{S})\simeq  \tau \iint\limits_{-\infty}^{\quad\infty}
[1-\Theta(z,\mathcal{S};\bs{x})][1-I_\mathcal{S}(\bs{x})]
d \bs{x}+ \tau
\iint\limits_{\mathcal{S}}[1-z\Theta(z,\mathcal{S};\bs{x})] d
\bs{x}\,,
\end{equation}
where $\Theta(z,\mathcal{S};\bs{x})$ is the solution of
Eq.~(\ref{7}) with (\ref{8}) or, equivalently, is the solution
of
\begin{equation}
\label{31}
\Theta=G\left[\Theta\otimes\phi-(1-z)
I_\mathcal{S}\Theta\otimes \phi\right]\,.
\end{equation}
where the function $G$ is defined in (\ref{12}).

\subsection{Factorization procedure \label{factoel}}

To find a reasonable approximate expression for the sought GPF
$\Theta(z,\mathcal{S};\bs{x})$, notice that if $\ell\gg d$
(or if $n$ is close to $1$) then the characteristic spatial
scale associated with the GPF
$\Theta(z,\mathcal{S};\bs{x})$ becomes greater than
$d$. Therefore, without essential error,
one may replace $\Theta\otimes\phi$ by
$\Theta$ in (\ref{31}). In addition, we take into account
the finiteness of the domain $\mathcal{S}$ by using the factorization
procedure of replacing the last term of the argument of the
function $G$ in (\ref{31}) as follows:
\be
I_\mathcal{S}(\bs{x})\Theta(z,\mathcal{S};\bs{x})\otimes
\phi(\bs{x})\simeq \Theta(z,\mathcal{S};\bs{x})~
p_\mathcal{S}(\bs{x})\,,
\label{mgkmss}
\ee
where $p_\mathcal{S}(\bs{x})$ remains to be specified. We will show below
that $p_\mathcal{S}(\bs{x})$ may be interpreted as the overall
fraction of aftershocks, triggered by a mother earthquake at position $\bs{x}$,
which fall within the domain $\mathcal{S}$.
The factorization procedure
amounts to replacing a convolution integral by an algebraic term.
This factorization approximation is a crucial step of our analysis and
will be justified further below.
As a result of its use, the nonlinear integral equation (\ref{31})
transforms into the functional equation
\begin{equation}\label{32}
\Theta=G[(1+(z-1) p_\mathcal{S}(\bs{x}))\Theta]\,.
\end{equation}
It is easy to show that, if relation (\ref{32}) holds,
then the average of the number of aftershocks corresponding to it
is equal to
\begin{equation}\label{33}
\langle R\rangle= \frac{n}{1-n}\, p_\mathcal{S}\,.
\end{equation}

In the next subsection, we shall clarify what is the probabilistic sense of the
parameter $p_\mathcal{S}$. Here, it is sufficient to remark that one can
determine it from a consistency condition: choose
$p_\mathcal{S}(\bs{x})$ such
that the r.h.s. of Eq.~(\ref{33}) is equal to the true
$\langle R(\mathcal{S};\bs{x})\rangle$. This gives
\begin{equation}
\label{34}
p_\mathcal{S}(\bs{x})= \frac{1-n}{n} \langle
R(\mathcal{S};\bs{x})\rangle\,, \qquad
\tilde{p}_\mathcal{S}(\bs{q})=\tilde{I}_\mathcal{S}(\bs{q})
\tilde{\phi}(\bs{q}) \frac{1-n
}{1-n\tilde{\phi}(\bs{q})}\,.
\end{equation}
Figure~6 plots $p_\mathcal{S}(\bs{x})$ defined by expression (\ref{34})
as a function of dimensionless distance $x/\ell$.

One can observe that, for $\ell\gg d$, the factor
$p_\mathcal{S}(\bs{x})$ approaches a rectangular function.
We can use this observation to help determine the statistics of the
number of events in a finite space-time window, using the
approximation $p_\mathcal{S}(\bs{x})\simeq \text{const}=p$
for $\bs{x}\in \mathcal{S}$. We define the parameter $p$
as the space average of $p_\mathcal{S}(\bs{x})$ over the window's
area $\mathcal{S}$:
\begin{equation}
\label{35}
p\simeq \frac{1}{S} \iint\limits_{\mathcal{S}}
p_\mathcal{S}(\bs{x}) d \bs{x}\,.
\end{equation}
$p$ is thus the average over all possible spatial positions of 
mother earthquakes of the fraction of aftershocks which fall within
the space-time window $\mathcal{S}$.
The approximation  $p_\mathcal{S}(\bs{x})\simeq \text{const}=p$
for $\bs{x}\in \mathcal{S}$ allows us to simplify the last term of (\ref{30})
as follows:
\be\tau
\iint\limits_{\mathcal{S}}[1-z\Theta(z,\mathcal{S};\bs{x})]
d \bs{x} \simeq \tau S [1-z\Theta(z;p)]~,
\ee
where $\Theta(z,p)$ is the solution of
\begin{equation}\label{36}
\Theta(z;p)= G[(1+(z-1)p)\Theta(z;p)]\,,
\end{equation}
which is derived from equation (\ref{32}).

Complementarily, as can be seen from figure~6,
$p_\mathcal{S}(\bs{x})$ is small outside the window space domain
$\mathcal{S}$. It implies that, outside
$\mathcal{S}$, one may replace Eq.~(\ref{32}) by its
linearized version. As a result, we get
\be
1-\Theta(z,\mathcal{S};\bs{x})\simeq \frac{n}{1-n} (1-z)
p_\mathcal{S}(x)\,.
\label{mgmsmwwkl}
\ee
Therefore, the first term in the r.h.s. of
Eq.~(\ref{30}) transforms into
\be
\tau \iint\limits_{-\infty}^{\quad\infty}
[1-\Theta(z,\mathcal{S};\bs{x})][1-I_\mathcal{S}(\bs{x})]
d \bs{x}\simeq q \tau S\, \frac{n}{1-n} (1-z)~,
\ee
where
\be
q= \frac{1}{S} \iint\limits_{-\infty}^{\quad\infty}
p_\mathcal{S}(\bs{x})[1-I_\mathcal{S}(\bs{x})]d\bs{x}\,.
\ee
Taking into account that, due to (\ref{34}),
\be
\iint\limits_{-\infty}^{\quad\infty}
p_\mathcal{S}(\bs{x})d\bs{x}= S\,,
\ee
we obtain
\be
q \simeq 1-p~.
\ee
Putting all these
approximations together allows us to rewrite Eq. (\ref{30}) in the form
\begin{equation}
\label{37}
L(z,\tau,\mathcal{S})\simeq \tau S \left[ \frac{n}{1-n}
(1-p) (1-z)+1-z \Theta(z;p)\right]\,.
\end{equation}
In what follows, we shall select a value of the parameter
$p$ which takes into account the finiteness of the window's
spatial domain $\mathcal{S}$, to better fit empirical data on the statistics
of seismic rates in finite space-time bins.

\subsection{Probabilistic meaning of the factorization approximation
(\ref{mgkmss}) leading to (\ref{32}) and (\ref{34})}

The factorization approximation (\ref{mgkmss}) has the
following implication. Calling $\Theta(z)$ the GPF of the total number of
aftershocks triggered over the whole space by
some earthquake source, one can then determine the corresponding GPF
$\Theta(z,\mathcal{S};\bs{x})$ taking into account
the finiteness of the space domain $\mathcal{S}$ by using the relation
\begin{equation}
\label{38}
\Theta(z,\mathcal{S};\bs{x})=\Theta(q_\mathcal{S}+z
p_\mathcal{S})\,, \qquad
q_\mathcal{S}(\bs{x})=1-p_\mathcal{S}(\bs{x})\,.
\end{equation}
This expression (\ref{38}) has the following interpretation. Let the 
above mentioned earthquake source triggers $r$ aftershocks. Then, 
the number of those aftershocks which fall into
the space domain $\mathcal{S}$, is equal to
\be
R_m(\mathcal{S};\bs{x}|r)= X_1+X_2+\dots +X_r\,,
\ee
where $\{X_1,\dots,X_r\}$ are mutually independent
random variables equal to $1$ with probability
$p_\mathcal{S}$ and $0$ with probability
$q_\mathcal{S}=1-p_\mathcal{S}$. Thus, $p_\mathcal{S}$ is the fraction
of the aftershocks which fall into the domain $\mathcal{S}$.

The corresponding expression (\ref{38}) can be interpreted as follows.
It gives the exact solution for the GPF of some specific space-time
branching process, such that the pdf $f(\bs{y};\bs{x})$ of the
space positions $\bs{y}$ of each aftershock is the same
for all aftershocks and depends only on the position
$\bs{x}$ of the earthquake source. For this problem, we have
\be
p_\mathcal{S}(\bs{x})=
\iint\limits_{-\infty}^{\quad\infty} f(\bs{y};\bs{x})
I_\mathcal{S}(\bs{y}) d\bs{y}\,.
\ee
In the general case, the relation (\ref{38}) offers the possibility,
at least semi-quantitatively, to describe the characteristic
features of the space-time branching processes, by using
the probability $p_\mathcal{S}$ as an
effective independent parameter of the theory.

Let us mention a few useful consequences of the relation
(\ref{38}). It implies that the probability that $r$
aftershocks fall into the spatial domain $\mathcal{S}$ is equal to
\begin{equation}
\label{39}
P(r,\mathcal{S};\bs{x})= \sum_{k=r}^\infty P(k)~
B(k,r,\mathcal{S};\bs{x})\,,
\end{equation}
where $P(k)$ is the probability that some earthquake source
triggers $k$ aftershocks and
\begin{equation}
\label{40}
B(k,r,\mathcal{S};\bs{x})= \binom{k}{r}\,
p_\mathcal{S}^r\, q_\mathcal{S}^{k-r}\,.
\end{equation}
This binomial probability $B(k,r,\mathcal{S};\bs{x})$ is nothing but the
conditional probability that, if the mother
earthquake triggers $k\geqslant r$ aftershocks then, $r$
of them will fall into the spatial domain $\mathcal{S}$. If $r\gg
1$, expression (\ref{40}) can be approximated by its well-known
Gaussian asymptotics
\begin{equation}
\label{41}
B(k,r,\mathcal{S};\bs{x})= \frac{1}{\sqrt{2\pi k
p_\mathcal{S}\, q_\mathcal{S}}}
\exp\left[-\frac{(r-k p_\mathcal{S})^2}{2k\,
p_\mathcal{S}\,q_\mathcal{S}}\right]\,.
\end{equation}
If, in addition, $P(k)$ decays slowly, for instance if it has
a power asymptotic for $k\to\infty$, then expressions
(\ref{39}) and (\ref{41}) imply the asymptotic
relation
\begin{equation}
\label{slope}
P(r,\mathcal{S};\bs{x})\sim \frac{1}{p_\mathcal{S}}\,
P\left(\frac{r}{p_\mathcal{S}}\right)\qquad
(r\to\infty)\,.
\end{equation}
In view of this last relation (\ref{slope}), it seems reasonable
to assume that the
asymptotic behavior of the probabilities of the number of events for $r\gg
1$ are the same for the case of a finite
$\mathcal{S}$ ($p_\mathcal{S}<1$) and for an unbounded one
($p_\mathcal{S}=1$).

\subsection{Large space-time windows}

In order to get more insight into the properties of the statistics
of seismic rates in finite space-time windows, it is useful to
study the statistics of seismic rates in the limit where the space and
time windows are large. In this case,
$\mathcal{N}_{\text{in}}$ in (\ref{29}) and $p$
in (\ref{35}) are both close to $1$ and one may replace
Eq.~(\ref{37}) by
\be
L(z,\tau,\mathcal{S})\simeq \tau S[1-z \Theta(z)]\,,
\ee
where $\Theta(z)$ is solution of the functional equation
\begin{equation}
\label{42}
\Theta(z)=G[z\Theta(z)]\,.
\end{equation}
Accordingly, the GPF of the number of events in a (large) space-time window
as given by (\ref{15}) takes the form
\be
\Theta_{\text{sp}}(z,\rho)=\hat{f}(\rho\, [1-z
\Theta(z)])\,.
\ee
Here and everywhere below,
\be
\rho=\langle\varrho\rangle\,\tau S~.
\label{gmksk}
\ee

Knowing the GPF $\Theta_{\text{sp}}(z,\tau;\mathcal{S})$,
the probability $P_{\text{sp}}(r;\rho)$ of
event numbers $r$ is obtained from the formula
\be
P_{\text{sp}}(r;\rho)=\frac{1}{r!}\left.
\frac{\partial^r \hat{f}(\rho\,[1-z
\Theta(z)])}{\partial z^r}\right|_{z=0}~.
\label{mgjghirik}
\ee
Equivalently, the integral representation of (\ref{mgjghirik}) reads
\begin{equation}
\label{43}
P_{\text{sp}}(r;\rho)= \frac{1}{2\pi i}
\oint\limits_{\mathcal{C}} \hat{f}(\rho\, [1-z
\Theta(z)]) \frac{dz}{z^{r+1}}\,,
\end{equation}
where $\mathcal{C}$ is a sufficiently small contour in the
complex plane $z$ around the origin $z=0$.

The main difficulty in calculating
$P_{\text{sp}}(r;\rho)$ comes from the fact that the GPF
$\Theta(z)$ is defined only implicitly by Eq.~(\ref{42}).
To overcome this difficulty, we rewrite the
integral in (\ref{43}) in the following equivalent form
\be
P_{\text{sp}}(r;\rho)= \frac{1}{2\pi i
r}\oint\limits_{\mathcal{C}} \frac{d\hat{f}(\rho\, [1-z
\Theta(z)])}{z^r} \qquad (r>0)
\label{mggkghklgf}
\ee
and use the new integration variable $y=z \Theta(z)$.
Expression (\ref{42}) shows that the inverse function of $y$ is $z=y/G(y)$. As
a result, the equation (\ref{mggkghklgf}) transforms into
\begin{equation}
\label{44}
P_{\text{sp}}(r;\rho)= \frac{\rho}{2\pi i
r}\oint\limits_{\mathcal{C}'}G^r(y) Q(y;\rho)
\frac{dy}{y^r}\,,
\end{equation}
where
\begin{equation}
\label{45}
Q(z;\rho)= \frac{1}{\rho}  \frac{d \hat{f}[\rho\,
(1-z)]}{dz}
\end{equation}
and $\mathcal{C}'$ is a contour enveloping the origin
$y=0$ in the complex plane $y$. One may interpret
$Q(z;\rho)$ (\ref{45}) as the GPF of some random integer
$R_\rho$ such that $\langle R_\rho\rangle=\rho$.

Notice that Eq.~(\ref{44}) has a simple probabilistic
interpretation. Indeed, it follows from (\ref{44}) that
\begin{equation}\label{46}
P_{\text{sp}}(r;\rho)= \frac{\rho}{r}\, \text{Pr\,}
\{R_\rho+ R(r)= r-1\}\,,
\end{equation}
where $R_\rho$ is a random integer with GPF
$Q(z;\rho)$ given by (\ref{45}) while
\begin{equation}
\label{47}
R(r)=R_1+R_2+\dots +R_r~,
\end{equation}
where $\{R_1,R_2,\dots,R_r,\dots\}$ are mutually
independent random integers with GPF $G(z)$. This
implies that the probabilities of each such random
variable $R_i$, $i=1, ..., r$, has the power law
asymptotics (\ref{14}).

This remark provides a simple analysis of the asymptotic
behavior of the probabilities $P_{\text{sp}}(r;\rho)$
for $r\gg 1$, by using expression (\ref{46}). For
$1<\gamma<2$, the asymptotics of the probability
$P(k;r)$ that the sum (\ref{47}) is equal to $k$ goes
to, for large $r$,
\begin{equation}
\label{48}
P(k|r)\simeq \frac{1}{(\epsilon r)^{1/\gamma}}
\psi_\gamma\left( \frac{k-n r}{(\epsilon
r)^{1/\gamma}}\right)\,,
\end{equation}
where
\be
\epsilon= -\left(n \frac{\gamma-1}{\gamma}\right)^\gamma
\Gamma(1-\gamma)
\ee
and $\psi_\gamma(x)$ is the stable L\'evy distribution with the
two-sides Laplace transform
\be
\int_{-\infty}^\infty \psi_\delta(x) e^{-sx}\, dx= e^{s^\delta}\,.
\ee
It is known that
\be
\psi_\delta(x)\sim
\frac{x^{-\delta-1}}{\Gamma(-\delta)}\qquad
(x\to\infty)\,, \qquad \psi_\delta(0)= \frac{1}{\delta
\Gamma(1-\frac{1}{\delta})}\,.
\ee
One can calculate $\psi_\delta(x)$ for any $1<\delta<2$,
using, for instance, the following integral
representation
\be
\psi_\delta(x)= \frac{1}{\pi}\int_0^\infty
\exp\left[-u^\delta+ux\cos\left(\frac{\pi}{\delta}\right)
\right]\sin\left[ux\sin\left(\frac{\pi}{\delta}\right)+
\frac{\pi}{\delta}\right]\,du\,.
\ee
For some numerical illustrations, we will use the case
$\delta=3/2$ for which the following analytical expression
is available
\be
\psi_{3/2}(x)=\tfrac{1}{\pi \sqrt{3}} \left[
\Gamma\left(\tfrac{2}{3}\right)
\mathstrut_1\!F_1\left(\tfrac{5}{6},\tfrac{2}{3},
\tfrac{4 x^3}{27}\right)-x\,
\Gamma\left(\tfrac{4}{3}\right)
\mathstrut_1\!F_1\left(\tfrac{7}{6},\tfrac{4}{3},
\tfrac{4 x^3}{27}\right)\right]\,.
\ee

For $r\gg \rho$, one can neglect the random integer $R_\rho$
in the r.h.s. of Eq.~(\ref{46}) and obtain from
(\ref{46}) and (\ref{48}) the following asymptotic
formula
\begin{equation}
\label{49}
P_{\text{sp}}(r;\rho)\simeq \frac{\rho}{r(\epsilon
r)^{1/\gamma}} \psi_\gamma\left( \frac{(1-n)
r-1}{(\epsilon r)^{1/\gamma}}\right) \qquad (r\gg \rho)\,.
\end{equation}
If $1-n\ll 1$ (the branching is close to but not exactly critical),
Eq.~(\ref{49}) predicts the existence of two
characteristic power laws in the dependence of the
probabilities $P_{\text{sp}}(r;\rho)$ with $r$, a result
already derived in \cite{Saichevetal04}.
\begin{enumerate}
\item For
\be
1\ll r\ll r_*\,, \qquad{\rm with}~~ r_*=
\left(\frac{1}{1-n}\right)^{\gamma/(\gamma-1)}
\epsilon^{1/(\gamma-1)}
\ee
then,
\begin{equation}
\label{50}
P_{\text{sp}}(r;\rho)\sim r^{-1-1/\gamma}\,.
\end{equation}

\item For
\be
r\gg r_*~,
\ee
we recover the original power law
(\ref{14}) of the number of first generation aftershocks
\begin{equation}
\label{51}
P_{\text{sp}}(r;\rho)\sim r^{-1-\gamma}\,.
\end{equation}
\end{enumerate}
For values of the parameters
$\gamma=b/\alpha$ and $n$ which are typical of real seismicity
modeled by aftershock triggering processes, the cross-over number
$r_*$ separating
the two power laws can be very large. For instance, for
$\gamma=1.25$ and $n=0.9$, we obtain $r_*\simeq 10^4$.

\subsection{Prediction of the distribution of event numbers for large
time windows}

Starting from the general expression (\ref{15}) of the GPF
$\Theta_{\text{sp}}(z,\tau;\mathcal{S})$ with the approximation (\ref{37})
for $L(z,\tau,\mathcal{S})$ and using the relationship between the probability
$P_{\text{sp}}(r;\rho,p)$ and its GPF similar to
expression (\ref{mgjghirik}) and its
integral representation similar to (\ref{43}), we obtain the following
expression valid in the limit of large time windows
\be
P_{\text{sp}}(r;\rho,p)= \frac{1}{2\pi i}
\oint\limits_{\mathcal{C}} \hat{f}\left(\rho\, \left[
\frac{n}{1-n} (1-p) (1-z)+1-z \Theta(z;p)\right]\right)
\frac{dz}{z^{r+1}}
\label{mgm,mslsd}
\ee
where $\Theta(z;p)$ is the solution of the functional equation
(\ref{36}). Similarly to the change of variable used to go from
(\ref{mggkghklgf}) to (\ref{44}), we introduce the
new integration variable
\be
y= (1+(z-1)p)\Theta(z;p)\quad \iff\quad z=Z(y)=
\frac{1}{p} \left(\frac{y}{G(y)}+p-1\right)\,.
\ee
By construction of $y$, $\Theta(z;p)=G(y)$ which allows us to obtain the
following explicit expression
\be
\begin{array}{c}
P_{\text{sp}}(r;\rho,p)= \frac{1}{2\pi i} \times \\[4mm] \displaystyle
\oint\limits_{\mathcal{C}'} \hat{f}\left(\rho\, \left[
\frac{n}{1-n} (1-p) (1-Z(y))+1-Z(y) G(y)\right]\right)
\frac{dZ(y)}{dy} \frac{dy}{Z^{r+1}(y)}\,.
\end{array}
\label{fjjalala}
\ee
This expression (\ref{fjjalala}) allows us to make a precise
quantitative prediction for the dependence of the distribution
$P_{\text{sp}}(r;\rho,p)$
of the number $r$ of earthquakes per space-time window as a function of
$r$, once the model parameters $n, \gamma, p, \delta$ and $\rho$ are
given. We note that
Pisarenko and Golubeva have shown that the distribution of numbers has the same
tail as the distribution of seismic rates \cite{PisGol}. Thus, for
the tails, our results
can be interpreted either as statements on the distribution of
realized numbers of
earthquakes or on the distribution of average seismic rates.

We now turn to a brief description of the data analysis and of their fits
with (\ref{fjjalala}).

\section{Empirical analysis and comparison with theory}

We use the Southern Californian earthquakes
catalog with revised magnitudes (available from the Southern California
Earthquake Center) as it is among the best one in terms of quality
and time span. Magnitudes $M_L$
are given with a resolution of $0.1$ from 1932 to 2003, in a region
from approximately $32^\circ$ to $37^\circ$N in latitude, and from $-114^\circ$
to $-122^\circ$ in longitude.
In order to maximize the size and quality of the data used for the
analysis (to improve
the statistical significance), we consider the sub-catalog
spanning the time interval $1994-2003$ for $M_L>1.5$, which contains
a total of $86,228$ earthquakes.
The completeness of this sub-catalog has been verified in the standard way in
\cite{ouisor} by computing the complementary cumulative
magnitude distribution for each year from $1994$ to $2003$ included. The
stability of the linear relationship of the logarithm of the number
as a function
of magnitude $M_L$ for $M_L>1.5$ is taken as a diagnostic of completeness.

The spatial domain are covered by square boxes of ($L=5$ km) $\times$ ($L=5$ km),
which gives us $16046$ spatial bins, with many of them being actually
empty. This size
is a compromise between conflicting requirements. On the one hand,
a smaller
spatial resolution should not be used due to the errors of localization of earthquake epicenters,
which are of this order. On the other hand, increasing $L$
decreases very fast ($\sim 1/L^2$) the total number of boxes with which
the distribution of the number of events can be constructed.
Four different sizes for the time window are considered: $dt=1$ day ($3652$
temporal bins), $dt=10$ days, $dt=100$ days
and $dt=1000$ days. Combining the space and time windows leads to
space-time windows or bins of size $L^2 \times dt$. For instance, for
$dt = 1$ day, there is a
total of $54669$ spatio-temporal bins with at least one event, with
$4298$ non-empty spatial bins.

Figures~7-10 plot the empirical probability density functions
$P_{\text{data}}(r)$
of the number $r$ of earthquakes in the space-time bins described above.
The straight line in Figs.~8-10 is the best fit with a pure power law
\be
P_{\text{data}}(r) \sim 1/r^{1+\mu}
\label{mfm,mfls}
\ee
over the range $10 \leq r \leq 100$. 
The estimation for $\mu$ is found stable across different
time windows, since the fitted values are
$\mu = 1.65$ for $dt=100$ days, $\mu = 1.75$ for $dt=10$ days and
$\mu = 1.60$ for $dt=1$ days. However, one can also observe at the
same time that the pdf
becomes more and more curved in the larger portion of the bulk as the size $dt$
of the time window is increased. This behavior can be explained by our theory
as we know describe.

First, all parameters are let free to adjust optimally. The parameters are
the branching ratio $n$ defined by (\ref{mgmlele}), the exponent
$\gamma$ defined in (\ref{aera}), the exponent $\delta$ defined in
(\ref{disrholl2}),
the parameter $p_\mathcal{S} \simeq p$ 
coming from the factorization procedure (\ref{mgkmss}) and given by
(\ref{34}) and (\ref{35}) (which is roughly equal to the overall 
fraction of aftershocks triggered by sources
in the domain $\mathcal{S}$ which fall within the same domain $\mathcal{S}$), and
the average number $\rho$ of spontaneous earthquake source per space-time
bin defined in (\ref{gmksk}). The theoretical curves shown in Figures~7-10
are obtained by a numerical integration of (\ref{fjjalala}) for the set
of parameters $n=0.96$, $\gamma = 1.1$, $p=0.25$, $\delta = 0.15$
and $\rho=0.0019~ dt$ days with $dt$ in units of days (thus equal
to $1000$ for Fig.7). Note that a given $\rho$ for a given space-time window
$[t,t+\tau]\times \mathcal{S}$
translates into the following average number of events inside that window:
$\langle R_{\text{sp}}(\tau,\mathcal{S})\rangle \simeq
\frac{\rho}{1-n}\simeq 0.02-0.16 \times dt ~(\text{in units of
days})$, taking the mean value $n=0.96$.
These parameters are chosen to best fit
the empirical pdf for the largest time window $dt=1000$ days shown in
Fig.~7. They are
then frozen and the theoretical distribution is recalculated with
formula (\ref{fjjalala}) with no adjustable
parameters for the three other cases $dt=100$ days (Fig.~8), $dt=10$
days (Fig.~9)
and $dt=1$ day (Fig.~10). The theory is thus able to account simultaneously for
all the considered time windows, with no adjustable parameters for the three
smallest time windows.

Second, we test for the sensibility of the parameters.
We actually obtain practically the same quality of fits
for all four values $dt=1, ..., 1000$ days by varying
$\delta$ and $\gamma$ in the ranges $0.1 \leq  \delta <
0.2$ and $1 \leq \gamma  \leq 1.5$, by properly adjusting
the other parameters adaptively. We do not show the
corresponding theoretical curves as they are essentially
equally good to fit the data within the empirical noise.
The choice $\delta$ close to zero is consistent
with the choice of the Cauchy distribution as a proxy
for the heterogeneity of the spatial distribution of
spontaneous earthquake sources inferred for the stress
field and deduced from previous theoretical
\cite{Zolo1,Zolo2} and empirical analysis of earthquake
sources \cite{Kaganstress}.
The strong sensitivity of the fits with
respect to the fractal structure of the
spontaneous sources quantified by the exponent $\delta$
is a surprising but positive return of this work. We did not
expect a priori that the distribution of seismic rates
would teach so much about the heterogeneity of the seismic
active regions.
But there is a lack of sensibility with respect $\gamma$. Thus,
the distribution of seismic rates cannot be used to
constrain the productivity parameter $\alpha$
(through $\gamma= b/\alpha$) other than by
confirming the range of previous estimations of its value:
 $\alpha \approx 1$ \cite{Felzeretal02,HelmKaganJackson04}, $0.5
\leq \alpha < 1$ \cite{Consoleetal03,alpha,Zhuangetal}.
The sensitivity of the fits with respect to the branching ratio $n$ and
to the overall fraction $p$ of aftershocks which fall
within the domain $\mathcal{S}$ requires
a special discussion which will be reported elsewhere.

We have also used functional forms for $f_\delta(x)$ other than
(\ref{disrholl2}) to describe the pre-existing heterogeneity of
spontaneous earthquake sources, such as half-Gaussian, exponential as
well as different variants of power laws. Overall, we find that we need
$f_\delta(x)$ to have a power law tail close to the Cauchy distribution
in order to get a reasonable fit for all four time windows. This shows
that the ETAS model as well as any other model of this class of
triggered seismicity need to be generalized to account for a
pre-existing heterogeneity of the crust, which controls the occurrence
of the spontaneous earthquake sources.

We would also like to stress that, according to our theory,
the value of the exponent $\mu \approx 1.6$ used
in (\ref{mfm,mfls}) to fit the tails of the distributions shown in
Figs.~8-10 is
describing a cross-over rather than a genuine asymptotic tail. Recall that
the distribution of the total number of aftershocks has two power law
regimes $\sim 1/r^{1+{1 \over \gamma}}$ for $r < r_* \simeq
1/(1-n)^{\gamma/(\gamma -1)}$
and $\sim 1/r^{1+ \gamma}$ for $r>r_*$ \cite{Saichevetal04}. The existence
of this cross-over together with
the concave shape of the distribution at small and intermediate values of $r$
combine to create an effective power law with an apparent exponent
$\mu \approx 1.6$
larger than the largest asymptotic exponent $\gamma$.
We have verified this to be the case in synthetically generated distributions
with genuine asymptotics exponent $\gamma=1.25$ for instance, which could be
well fitted by $\mu \approx 1.6$ over several decades. We note also that
Pisarenko and Golubeva \cite{PisGol}, with a different approach applied to much
larger spatial box sizes in California, Japan and Pamir-Tien Shan, have
reported an exponent $\mu<1$ which could perhaps be associated with
the intermediate asymptotics characterized by the exponent $1/\gamma
< 1$, found
in our analysis \cite{Saichevetal04}. By using data collapse
with varying spatial box sizes on a California catalog,
Corral finds that the distribution of seismic
rates exhibits a double power-law behavior with $\mu \approx 0$ for small rates
and $\mu \approx 1.2$ for large rates \cite{Corral}.  The first regime
might be associated with the non universal bulk par of the distribution
found in our analysis. The second regime is perhaps compatible with the
prediction for the asymptotic exponent $\mu = \gamma$.

\section{Theoretical tests of the theory using
statistics conditioned on generation number \label{exmmflaa}}

In our theoretical development to obtain the prediction (\ref{fjjalala})
that could be compared with empirical data, we have been obliged
to make two approximations: (1) assuming that the duration $\tau$
of the time window $[t, t+\tau]$
is sufficiently large (i.e., the inequality (\ref{24}) holds), we have replaced
the GPF $\Theta(z,\tau,\mathcal{S};\bs{x})$ by its asymptotics
$\Theta(z,\mathcal{S};\bs{x})$; (2)
we have used a factorization procedure to take into
account quantitatively the finiteness of the spatial window
$\mathcal{S}$.

In this section, we attempt to clarify further the domain of
application of these two approximations by testing them
on other event statistics conditioned on fixed number of generations.
Numerical calculations of the exact PDF are compared with
the approximations.

\subsection{Large spatial windows}

Let us consider the statistics of aftershocks
triggered over the whole space during the first $k$
generations by some mother event. The corresponding GPF of
the number of aftershocks triggered in the course of $k$
generations is defined by the following iterative recurrence equation
\begin{equation}
\label{52}
\Theta_k(z)=G[z\Theta_{k-1}(z)]\,, \qquad
\Theta_1(z)=G(z)\,.
\end{equation}
One can calculate the corresponding probabilities of
aftershock numbers by using the Cauchy integral
\begin{equation}\label{53}
P_k(r)= \frac{1}{2\pi} \oint\limits_{\mathcal{C}}
\Theta_k(z) \frac{dz}{z^{r+1}}~.
\end{equation}
Furthermore, we can make use of the knowledge that,
as $k\to\infty$, the GPF
$\Theta_k(z)$ converges to the asymptotic GPF $\Theta(z)$
which is the solution of Eq.~(\ref{42}). It is easy
to show that one can calculate the corresponding
probabilities using an equality analogous to (\ref{44}):
\begin{equation}
\label{54}
P(r)= \frac{1}{2\pi i(r+1)} \oint\limits_{\mathcal{C}'}
G^{r+1}(y) \frac{dy}{y^{r+1}}\,.
\end{equation}
Figure~11 shows the distribution $P_k(r)$ of
aftershocks numbers, obtained by a numerical calculation
of the integral (\ref{53}) for different values $k=1, 2, 3, 5, 8$
and for $\gamma=1.25$ and in the critical case $n=1$.
It also shows the corresponding asymptotic distribution for $k \to +\infty$
obtained by integration of (\ref{54}).
Note that, even for in this critical case, the distribution for $k=8$
generations is already almost undistinguishable from the asymptotic
distribution including an infinite number of generations, at least
for $r\leqslant 250$. Figure~12 clarifies further the convergence rate
by plotting the ratio
\begin{equation}\label{55}
p_k(r)= \frac{P_k(r)}{P(r)}
\end{equation}
for different values $k$.

The information on the number of generations necessary to reach
the asymptotic regime gives us the possibility of
estimating the corresponding characteristic time beyond which
the asymptotic distribution $P(r)$ becomes a
good approximation of $P_k(r)$. Let $T(k)$ denote the random time
at which a $k$-th generation aftershock is triggered. It is equal to
\begin{equation}\label{56}
T(k)=\tau_1+\tau_2+\dots +\tau_k\,,
\end{equation}
where $\{\tau_1,\tau_2,\dots\tau_k, \dots\}$ are
mutually independent random waiting times between the occurrence of
a mother earthquake and one of its first-generation aftershock.
We define the $\omega$-th waiting time quantile $t(\omega,k)$ of
generation $k$ by
\begin{equation}
\label{57}
\mathcal{Q}[t(\omega,k),k]=\text{Pr}\{T(k)>t(\omega,k)\}=\omega~.
\end{equation}
Thus, $1-\omega$ is the probability that the
duration of any chain of $k$ successive generations of triggered
aftershocks is smaller than $t(\omega,k)$. Choosing some
confidence level (for example $1-\omega=0.9$), one may
assert that, during the time $t(\omega,k)$, all aftershocks of the $k$-th
generation have already been triggered. Let
$k=k_*$ be such that the corresponding probability
$P_{k_*}(r)$ is close to the asymptotic $P(r)$. Then, one
may interpret
\be
t_*=t(\omega,k_*)
\label{mkhklf}
\ee
as an estimation of the characteristic time for the validity of the asymptotic
distribution $P(r)$.

The asymptotic expression for the probability
$\mathcal{Q}(t,k)$ defined by (\ref{57}) for $k\gg 1$ can be
determined by using the fact that the terms $\tau_k$ of the sum
(\ref{56}) are determined by Omori's law (\ref{Omori}) with
$0<\theta<1$. For
$k\gg1$, $\mathcal{Q}(t,k)$ is asymptotically close to
\begin{equation}
\label{58}
\mathcal{Q}(t,k)= F_\theta\left(
\frac{t}{c[k\Gamma(1-\theta)]^{1/\theta}}\right)~,
\end{equation}
where
\be
F_\theta(x)=\int_x^\infty \varphi_\theta(y)\, dy
\ee
and $\varphi_\theta(x)$ is the one-sided L\'evy stable
distribution defined by the Laplace transform
\be
\hat{\varphi}_\theta(u)=\int_0^\infty \varphi_\theta(x)
e^{-ux}\, dx= e^{-u^\theta}\,.
\ee
In particular
\be
F_{1/2}(x)=
\text{erf\,}\left(\frac{1}{2\sqrt{x}}\right)\,.
\ee
The following asymptotic behavior holds
\begin{equation}
\label{59}
F_\theta(x)\simeq
\frac{x^{-\theta}}{\Gamma(1-\theta)}\qquad (x\gg 1)\,.
\end{equation}
Substituting (\ref{58}) and (\ref{59}) into (\ref{57}), we
obtain the following estimation for $t^*$ defined by (\ref{mkhklf})
\begin{equation}
\label{60}
t_*\simeq c \left(\frac{k_*}{\omega}\right)^{1/\theta}\,.
\end{equation}
For instance, for $\theta=1/2$, $k_*=8$, $\omega=0.1$ and
$c=2$ minutes, then $t_*\simeq 9$ days. Note that $t_*$ is
highly sensitive to the value of $\theta$. Indeed, $\theta=1/3$ (resp. $2/3$)
with all other parameters being the same gives $t_*\simeq 700$ days
(resp. $t_*\simeq 1$ day).

Expression (\ref{60}) is related to condition (\ref{24})
(and actually improves on it) as follows. The condition
(\ref{24}) with (\ref{23}) can be expressed by
introducing, similarly to the reasoning leading to
(\ref{60}), some small threshold $\omega \ll 1$ such
that (\ref{24}) translates into
$\mathcal{N}_{\text{out}}(\tau) \lesssim \omega$.
Correspondingly, we can introduce $\tau^*$ such that, if
$\tau > \tau^*$, then the duration of the window is
large at the $\omega$ level:
\be 
\omega \simeq {n \over
\Gamma(1-\theta)} \left({c_1 \over
\tau^*}\right)^{\theta}~. 
\ee 
Using (\ref{ngjmss,s,}), we
get 
\be 
\tau^* \simeq c \left({n \over \omega
(1-n)}\right)^{1/\theta}~. \label{gmjmloels} 
\ee 
The two
expressions (\ref{gmjmloels}) and (\ref{60}) have a
similar structure. The only difference is that the
characteristic generation number $k_*$ is replaced by
the factor $n/(1-n)$. For $n$ not too close to $1$,
$n/(1-n)$ gives a not unreasonable estimation of $k_*$.
For $n$ close to $1$, expression (\ref{60}) should be
preferred as it provides an improvement to (\ref{24})
based on the calculation of quantiles rather than on the
mean rate behavior.

\subsection{Finite spatial windows}

A natural generalization of the iterative procedure (\ref{52}) for
finite spatial window $\mathcal{S}$ allows to estimate the
corresponding aftershock statistics. Consider an earthquake occurring at
point $\bs{x}$. Then, the pdf of the space positions $\bs{y}$ of
an aftershock of the $k$-th generation is given by
$\phi_k(\bs{y}-\bs{x})$, where
\be
\phi_k(\bs{y})=
\phi(\bs{y})\underbrace{\otimes\dots\otimes}_k \phi(\bs{y})
\ee
is the $k$-times convolution of the space propagator $\phi(\bs{y})$
(one example is
given by (\ref{9})). Correspondingly, the probability for an
aftershock of the $k$-th
generation to fall into the space window
$\mathcal{S}$ is equal to
\be
p_k(\mathcal{S};\bs{x})= \iint\limits_{\mathcal{S}}
\phi_k(\bs{y}-\bs{x}) d \bs{y}\,.
\label{mghjjslks}
\ee
Provided these probabilities are known, one can determine the GPF
$\Theta_k(z,\mathcal{S};\bs{x})$
of the number of aftershocks of the $k$ generation occurring in the
spatial domain
$\mathcal{S}$ by using the following iteration
\be
\begin{array}{c}
\Theta_1(z,\mathcal{S};\bs{x})=
G\left(p_1(z-1)+1\right)\,,\\
\Theta_2(z,\mathcal{S};\bs{x})=
G\left[\left(p_1(z-1)+1\right)
G\left(p_2(z-1)+1\right)\right]\,,\\
\Theta_2(z,\mathcal{S};\bs{x})=
G\left[\left(p_1(z-1)+1\right)
G\left[\left(p_2(z-1)+1\right)
G\left[\left(p_3(z-1)+1\right)\right]\right]\right]\,,
\end{array}
\ee
and so on up to the order $k$. Then, the distribution of the number
of aftershocks
of the $k$ generation is given by
\begin{equation}
\label{60a}
P_k(r,\mathcal{S};\bs{x})= \frac{1}{2\pi}
\oint\limits_{\mathcal{C}}
\Theta_k(z,\mathcal{S};\bs{x}) \frac{dz}{z^{r+1}}\,.
\end{equation}

These expressions are general and hold for any space propagator
$\phi(\bs{y})$. Let
us now specialize to the form (\ref{9}) for the spatial propagator
$\phi(\bs{y})$,
with $\eta=1$,
\be
\phi_k(\bs{x})= \frac{k d}{2\pi (x^2+ k^2 d^2)^{3/2}}~
\ee
corresponding to an asymptotic $1/|{\bf x}|^3$ decay law often argued
on the basis of the shape of the elastic Green function in a three
dimensional space.
 From (\ref{mghjjslks}), we then have
\begin{equation}
\label{61}
p_k(\mathcal{S};\bs{x})= \mathcal{P}\left(
\frac{\ell}{kd}, \frac{x}{kd}\right)
\end{equation}
where
\be
\mathcal{P}(u,v)=\frac{2}{\pi} \int_0^u
E\left(\frac{4vs}{1+(v+s)^2}\right) \frac{s
ds}{[1+(v-s)^2] \sqrt{1+(v+s)^2}}\,.
\ee
Here, $E(m)$ is the complete elliptic integral
\be
E(m)= \int_0^{\pi/2} \sqrt{1-m \sin^2 \epsilon}\,
d\epsilon\,.
\ee
Figure~13 shows the distributions (\ref{61}) for $\ell=10 d$ (recall
that $\ell$ is the radius of the assumed circular domain $\mathcal{S}$
centered on the origin, which has been used in (\ref{mghkhlr})) and for
different positions $\bs{x}$ of the mother earthquake, given by
$x/\ell=0;~0.4; ~0.6; ~0.8; ~1; ~1.2;~1.4;~1.6$ from top to bottom. The
separation by the curve for $x/\ell=1$ into two families has a simple
explanation. For $x/\ell \leq 1$, the mother event lies within the
spatial domain $\mathcal{S}$ of interest and it is thus counted as
generation $0$. Its immediate aftershocks are most probably adjacent to
it and thus have a large probability to also fall within $\mathcal{S}$.
As the number $k$ of generation increases, aftershocks diffuse away and
are less and less likely to fall within $\mathcal{S}$. In contrast, for
$x/\ell > 1$, the mother earthquake falls outside $\mathcal{S}$.
Therefore, there is not event at the zeroth generation in $\mathcal{S}$,
hence the curves start from zero. The first generations of aftershocks
which are most likely to be nearby the mother earthquake fall rarely
within $\mathcal{S}$. Only as aftershocks of higher generation levels
develop and diffuse away from the mother earthquake, can they invade
$\mathcal{S}$. Of course, at large generation numbers, the aftershocks
diffuse away from any finite spatial domain, explaining the decay of
$P_k(\mathcal{S};\bs{x})$ to zero for large $k$'s.

Figure~14 plots the asymptotic distribution $P(r,\mathcal{S};\bs{x})$
as a function of the number $r$ of aftershocks
for $\gamma=1.25$, $n=0.99$, for different values of the radius $\ell$ of the
disk $\mathcal{S}$. The mother earthquake is assumed to occur at the
origin, that is,
at the center of the disk $\mathcal{S}$.
$P(r,\mathcal{S};\bs{x})$ is obtained by using (\ref{60a}) for $k=25$
generations, which is certainly a very good approximation
to $P(r,\mathcal{S};\bs{x}) = P_{k \to +\infty}(r, \mathcal{S};\bs{x})$.
Figure~15 plots $P(r,\mathcal{S};\bs{x})$ as a function of the number
$r$ of aftershocks,
at fixed $\ell/d=10$ for various positions of the mother earthquake,
for the same parameters $\gamma=1.25$, $n=0.99$.
One can observe that, when the mother earthquake is inside the disk
$\mathcal{S}$
($x= 0.8 \ell;\, 0.6 \ell;\, 0.4 \ell;\, 0.2 \ell;\, 0$), the corresponding
distributions are close to each
other as predicted in Section \ref{factoel}. When the mother is
outside the disk $\mathcal{S}$
($x=1.4 \ell;\, 1.2 \ell;\, 1$), the distributions differ from
the previous case and are significantly smaller. This gives additional
support in favor of the linear approximation (\ref{mgmsmwwkl}),
which we used in Section \ref{factoel}. More precisely, these properties
result directly from the analysis of section \ref{factoel}, in
which we notice below equation (\ref{34}) that, for $\ell\gg d$, the factor
$p_\mathcal{S}(\bs{x})$ approaches a rectangular function, which
leads to the approximation $p_\mathcal{S}(\bs{x})\simeq \text{const}=p$
for $\bs{x}\in \mathcal{S}$. This leads to the natural assumption that the
GPF $\Theta(z,\mathcal{S};\bs{x})$ is almost the
same for all interior source positions $\bs{x} \in \mathcal{S}$.
This means in turn that the corresponding
distribution $P(r,\mathcal{S};\bs{x})$ should be
the same for all $\bs{x} \in \mathcal{S}$. This remarkable fact is
illustrated in figure 15 in which the curves for the interior source
probabilities
merge.

\subsection{Testing the factorization approach}

We can now test the factorization approximation developed in
section \ref{factoel} to take into account the finiteness of the
space window $\mathcal{S}$ by comparing it with the approach
in term of the statistics over successive generations of the previous
section. We thus compare the asymptotic distribution
$P(r,\mathcal{S},\bs{x}=0)$, obtained by calculating the
integral (\ref{60a}) for a large enough generation number $k$ ($k=25$
is found to be sufficient), with the factorization approximation
\begin{equation}
\label{fact}
P(r,p)= \frac{p^r}{2\pi r} \oint\limits_{\mathcal{C}'}
\frac{dG(y)}{dy} \frac{G^r(y)\, dy}{[y-(1-p)G(y)]^r} \,,
\end{equation}
with an appropriate value of the parameter $p = p_\mathcal{S}$, defined
as the fraction of the aftershocks which fall into the domain $\mathcal{S}$.
Figure~16 shows this comparison for four different values of
$\ell/d = 10; ~7.5; ~5; ~2.5$. The upper curve in each panel is
the distribution (\ref{54}) for an infinite domain $\ell/d = +\infty$,
as a reference. There is some discrepancy between
$P(r,\mathcal{S},\bs{x}=0)$ and $P(r,p)$ given by (\ref{fact}). The
main difference is that the true distribution
$P(r,\mathcal{S},\bs{x}=0)$ decays faster than the factorization approximation
$P(r,p)$ for large $r$. We recall that, due to
(\ref{slope}), the asymptotic behavior of the distribution $P(r,p)$ obtained
under the factorization approximation is the same as for an infinite domain
given by (\ref{54}). Nevertheless, Figure~16 shows that an appropriate
choice of the parameter $p = p_\mathcal{S}$ allows us,
at least semi-quantitatively, to take into account the finiteness of the area
$\mathcal{S}$.

\section{Discussion}

We have presented a general formulation in terms of generating functions
of the space-time organization of earthquake sequences, in the
framework of general branching processes. We have applied this approach
to the ETAS (Epidemic-Type Aftershock Sequence) model of triggered
seismicity. In view of the formidable difficulty in obtained exact
solutions to the nonlinear integral equations involving the generating functions,
we have developed several approximation schemes which have been tested
by comparison with exact numerical calculations. We have used
the corresponding predictions to fit the distribution of seismic rates
in four finite space-time windows in a California seismic catalog. The
space-time windows differ by their time interval going from $dt=1$
day to $dt=1000$ days.
The fits have been
found to account satisfactorily for the empirical observation.
In particular, we have adjusted the parameters of the theory
on the largest
time window $dt=1000$ days and have then used these frozen parameter
values in the theory
to calculate the distribution for the smaller time windows. This tests
the rigidity of the theory to account simultaneously for the distributions
at different time scales. In this process, we have found it necessary to
augment the ETAS model
by taking account of the pre-existing frozen heterogeneity of
spontaneous earthquake sources. We have discussed the physical
justification of this generalization
in terms of pre-existing stress and fault networks, which constrain
the form of the pre-existing heterogeneity.
Our findings have also important implications
to assess the quality of models developed to forecast future seismicity,
and suggest to re-examine current procedures
which assume Poisson statistics in the construction of likelihood scores.

{\bf Acknowledgments:} We thank warmly G. Ouillon for
help in the analysis of the data and in the
preparation of the corresponding figures.
This work is partially supported by NSF-EAR02-30429, and by
the Southern California Earthquake Center (SCEC)
SCEC is funded by NSF Cooperative Agreement EAR-0106924 and USGS Cooperative
Agreement 02HQAG0008. The SCEC contribution number for this paper is xxx.

{}

\clearpage

\clearpage
\begin{psfrags}
\psfrag{P}{$P_1(r)$} \psfrag{}{$$}
\psfrag{r}{$r$}
\psfrag{g=1.25}{$\gamma=1.25$}
\psfrag{g=3}{$\gamma=3$}
\begin{quote}
\centerline{
\resizebox{16cm}{!}{\includegraphics{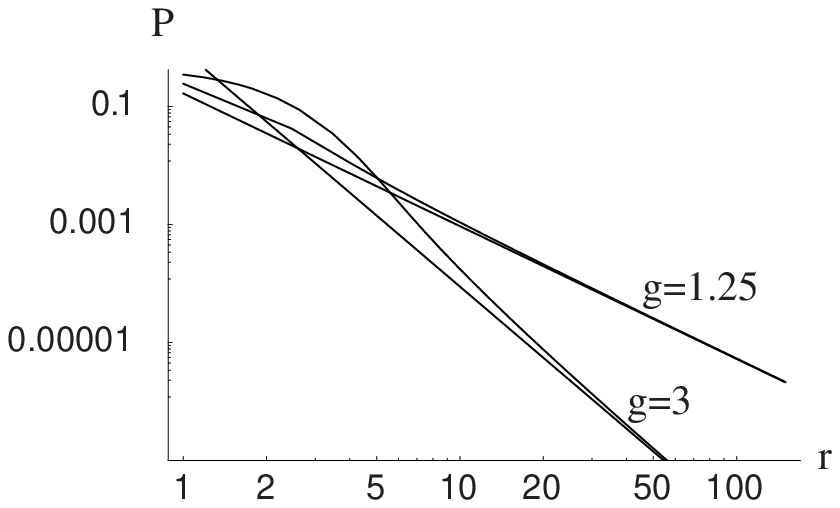}}} {\bf
Fig.~1:} \small{Plot of the probabilities (\ref{13}) and
their power law asymptotics (\ref{14}) for the infinite variance case
$\gamma=1.25$ and for a finite variance case $\gamma=3$. }
\end{quote}
\end{psfrags}

\clearpage
\begin{psfrags}
\psfrag{t'}{$t'$} \psfrag{S}{$\mathcal{S}$}
\psfrag{t}{$t$} \psfrag{t+t}{$t+\tau$}
\psfrag{x}{$\bs{x}$}
\begin{quote}
\centerline{
\resizebox{16cm}{!}{\includegraphics{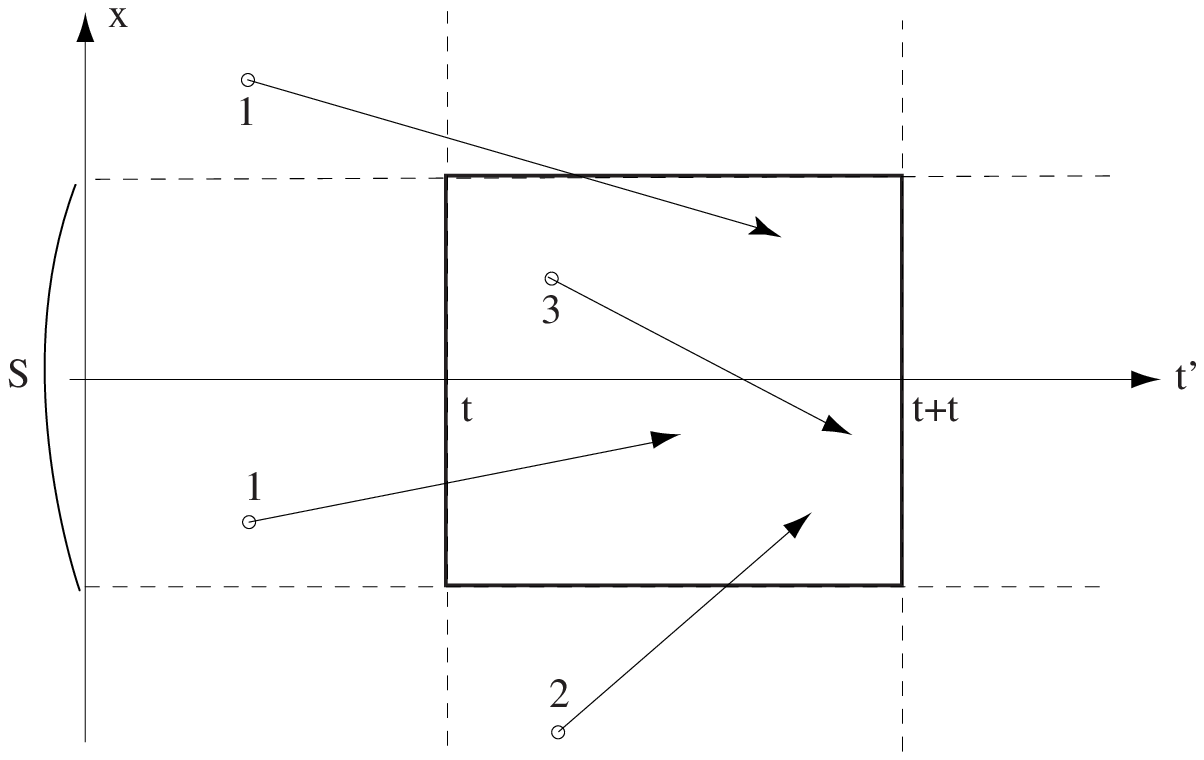}}}
{\bf Fig.~2:} \small{Illustration of the three different sets of
space-time locations for mother earthquakes contributing to
the three terms in the r.h.s. of expression (\ref{2}).}
\end{quote}
\end{psfrags}

\clearpage
\begin{psfrags}
\psfrag{N}{$\mathcal{N}_{\text{out}}(\tau)$}
\psfrag{t}{$\tau/c_1$}
\begin{quote}
\centerline{
\resizebox{16cm}{!}{\includegraphics{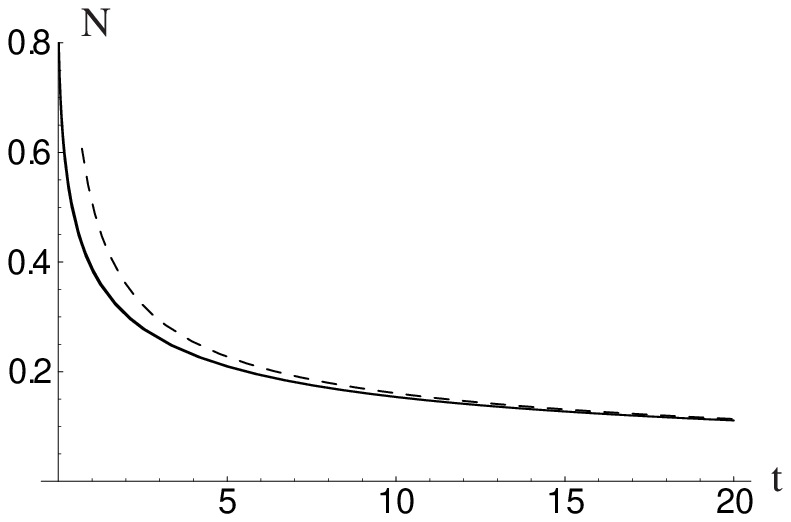}}}
{\bf Fig.~3:} \small{Plots of the exact rate
$\mathcal{N}_{\text{out}}(\tau)$, its fractional
approximation (\ref{21}) (which actually coincides with the exact
value) and its
asymptotic approximation (\ref{23}) obtained from (\ref{22})
(dashed line), for $n=0.9$ and $\theta=1/2$.}
\end{quote}
\end{psfrags}

\clearpage
\begin{psfrags}
\psfrag{R}{$\langle R(\tau,\mathcal{S};\bs{x})\rangle$}
\psfrag{x}{$x/d$}
\psfrag{t=c}{$\tau=c_1$}
\psfrag{t=inf}{$\tau=\infty$}
\begin{quote}
\centerline{
\resizebox{16cm}{!}{\includegraphics{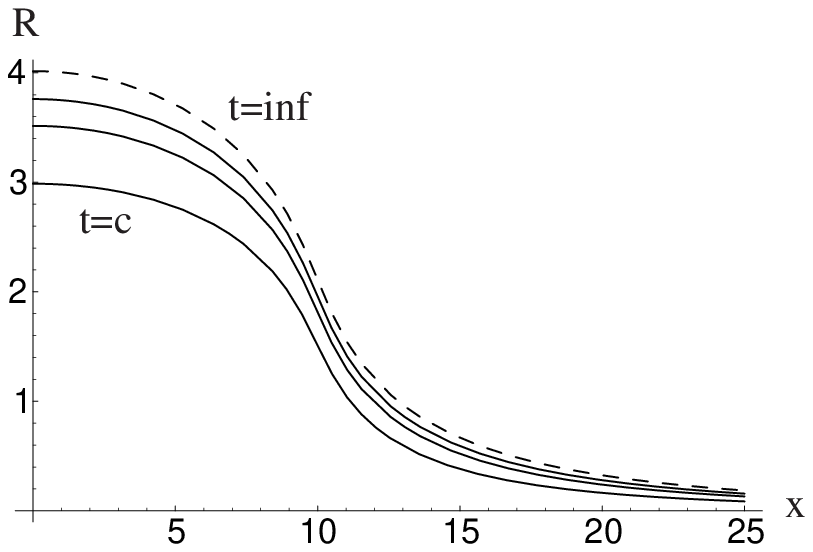}}} {\bf
Fig.~4:} \small{Plots of $\langle
R(\tau,\mathcal{S};\bs{x})\rangle$ for $\eta=1$,
$\ell=10\, d$, $\theta=1/2$ and for
$\tau=c_1;\,5\,c_1;\,20\,c_1$. Dashed line is the plot
of average of total number of aftershocks, falling into
the circle $\mathcal{S}$.}
\end{quote}
\end{psfrags}

\clearpage
\begin{psfrags}
\psfrag{N}{$\mathcal{N}_{\text{in}}(\mathcal{S})$}
\psfrag{l}{$\ell/d$}
\psfrag{h=1}{$\eta=1$}
\psfrag{h=2}{$\eta=2$} \psfrag{h=3}{$\eta=3$}
\begin{quote}
\centerline{
\resizebox{16cm}{!}{\includegraphics{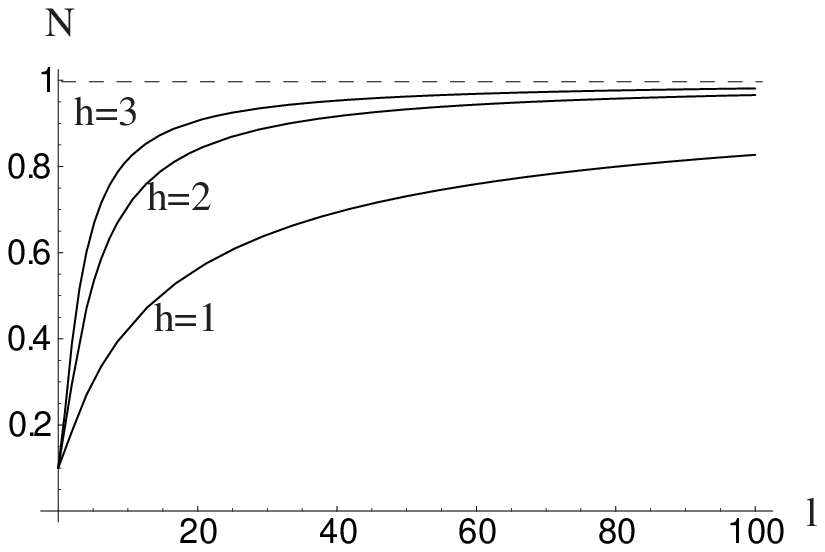}}} {\bf
Fig.~5:} \small{Plots of the relative rate
$\mathcal{N}_{\text{in}}(\mathcal{S})$ given by (\ref{29})
as a function of the dimensionless size
$\ell/d$ of the space domain, for different exponents $\eta=1;\,2;\,3$
of the space propagator. }
\end{quote}
\end{psfrags}

\clearpage
\begin{psfrags}
\psfrag{p}{$p_\mathcal{S}(x)$} \psfrag{x}{$x/\ell$}
\psfrag{n=0.9}{$n=0.9$} \psfrag{n=0.8}{$n=0.8$}
\psfrag{n=0.5}{$n=0.5$}
\begin{quote}
\centerline{
\resizebox{16cm}{!}{\includegraphics{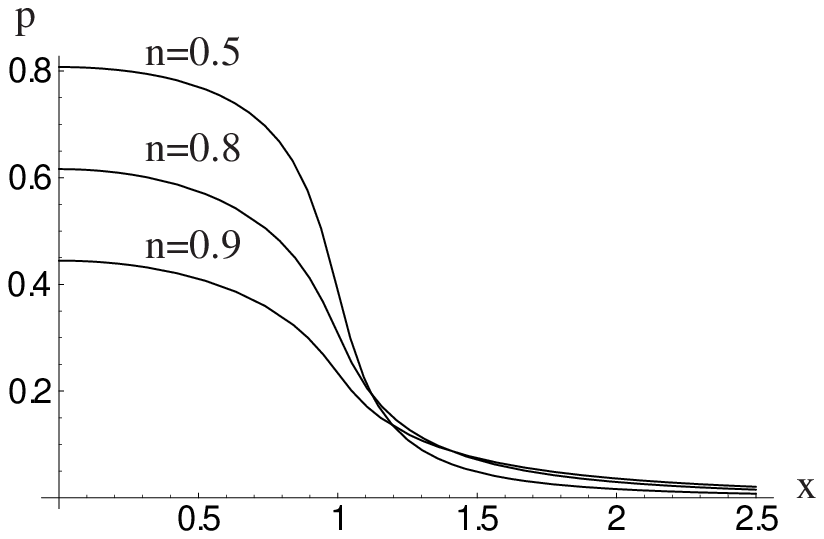}}}
{\bf Fig.~6:} \small{Plots of the self-consistent factor
$p_\mathcal{S}(\bs{x})$
defined by (\ref{34}) for $\ell=10\, d$ and for
$n=0.5;\,0.8;\,0.9$. }
\end{quote}
\end{psfrags}

\clearpage
\begin{quote}
\centerline{
\includegraphics[width=16cm]{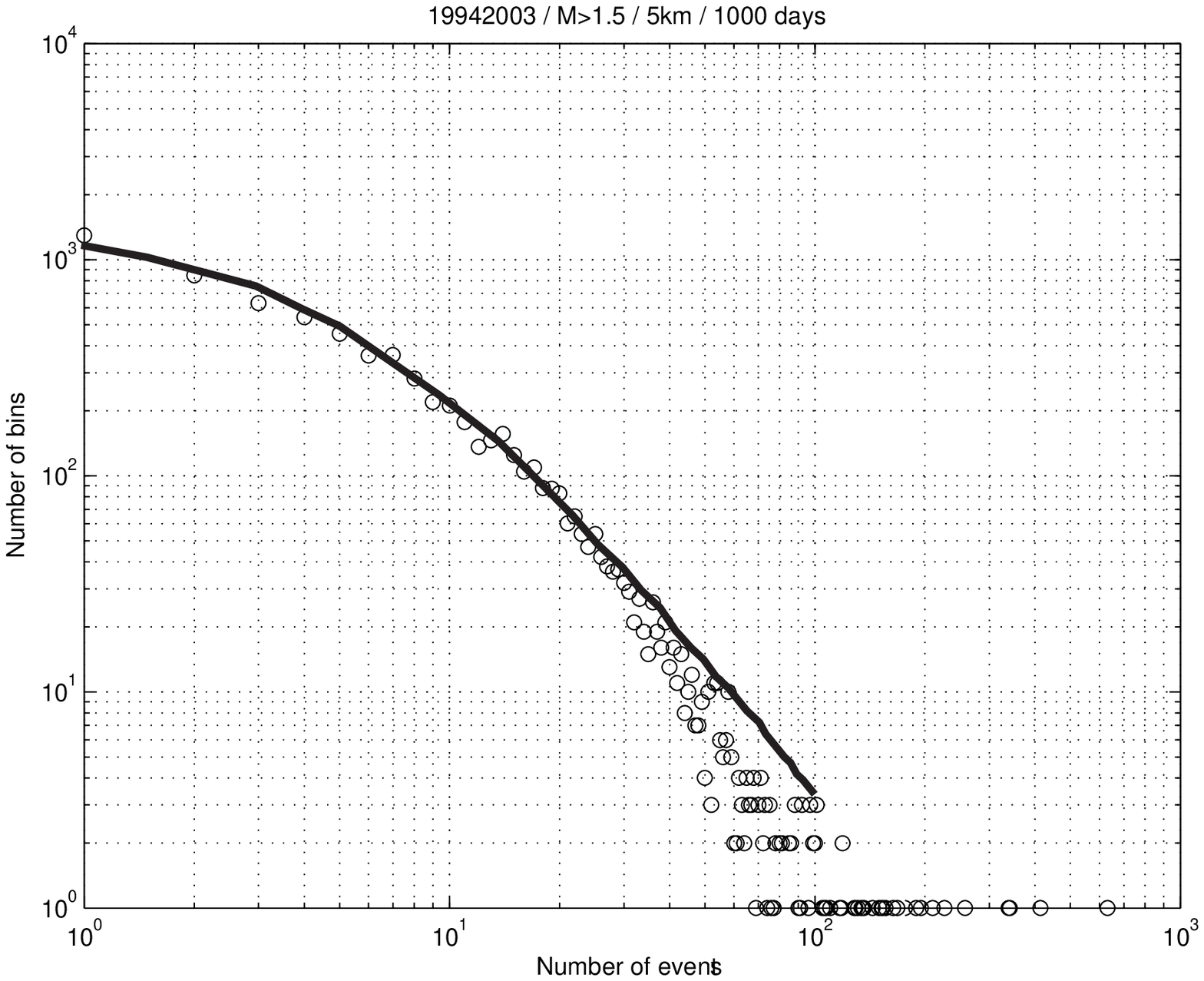}}
{\bf Fig.~7:} \small{Empirical
probability density functions $P_{\text{data}}(r)$
of the number $r$ of earthquakes in the space-time bins of size $5 \times 5$
km$^2$ and $dt=1000$ days. The continuous line is the fit of this data with
formula (\ref{fjjalala}) for the set of parameters $n=0.96$, $\gamma =
1.1$, $p_\mathcal{S}=0.25$, $\delta = 0.15$ and $\rho=0.0019$ km $\times
1000$ days. }
\end{quote}

\clearpage
\begin{quote}
\centerline{
\includegraphics[width=16cm]{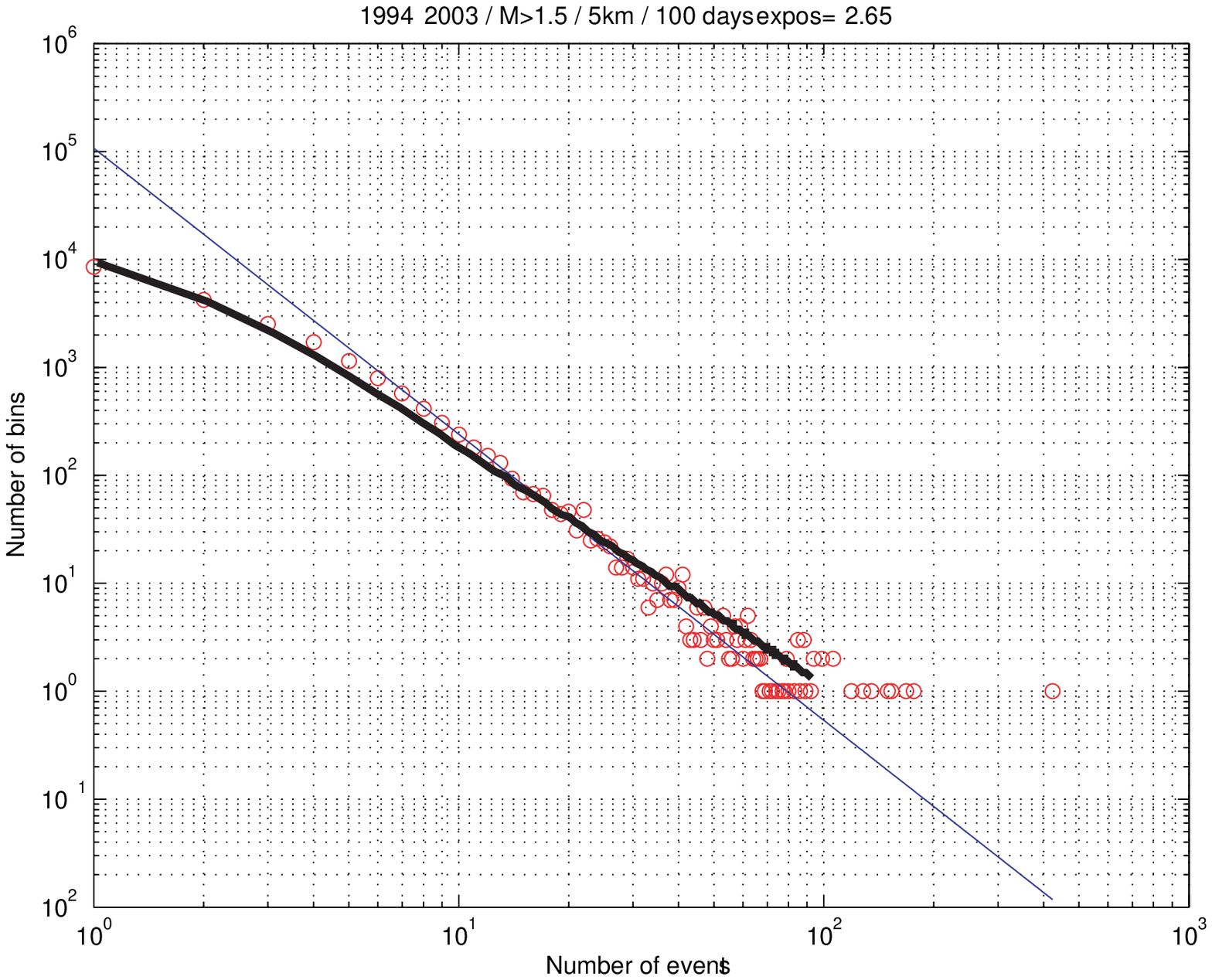}}
{\bf Fig.~8:} \small{Empirical
probability density functions $P_{\text{data}}(r)$
of the number $r$ of earthquakes in the space-time bins of size $5 \times 5$
km$^2$ and $dt=100$ days.
The straight line is the best fit with a pure power law
$P_{\text{data}}(r) \sim 1/r^{1+\mu}$ over the range $10 \leq r \leq
100$, which
gives $\mu = 1.65$. The continuous line is the theoretical prediction using
formula (\ref{fjjalala}) with $dt=100$ days and with no
adjustable parameters, as the parameters are fixed to the values adjusted
from the fit in Fig.7.}
\end{quote}

\clearpage
\begin{quote}
\centerline{
\includegraphics[width=16cm]{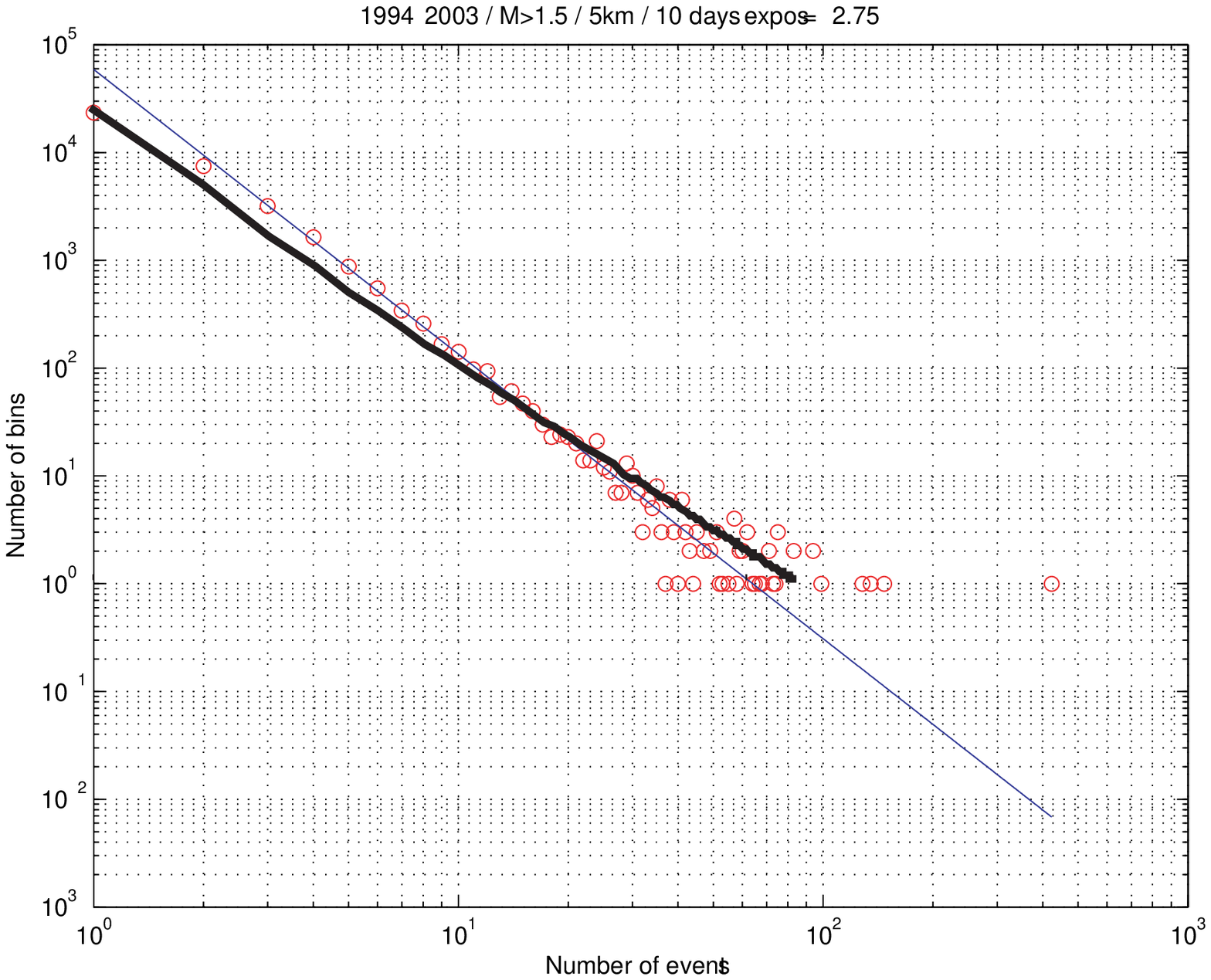}}
{\bf Fig.~9:} \small{Empirical
probability density functions $P_{\text{data}}(r)$
of the number $r$ of earthquakes in the space-time bins of size $5 \times 5$
km$^2$ and $dt=10$ days.
The straight line is the best fit with a pure power law
$P_{\text{data}}(r) \sim 1/r^{1+\mu}$ over the range $10 \leq r \leq
100$, which
gives $\mu = 1.75$. The continuous line is the theoretical prediction using
formula (\ref{fjjalala}) with $dt=10$ days and with no
adjustable parameters, as the parameters are fixed to the values adjusted
from the fit in Fig.7.}
\end{quote}

\clearpage
\begin{quote}
\centerline{
\includegraphics[width=16cm]{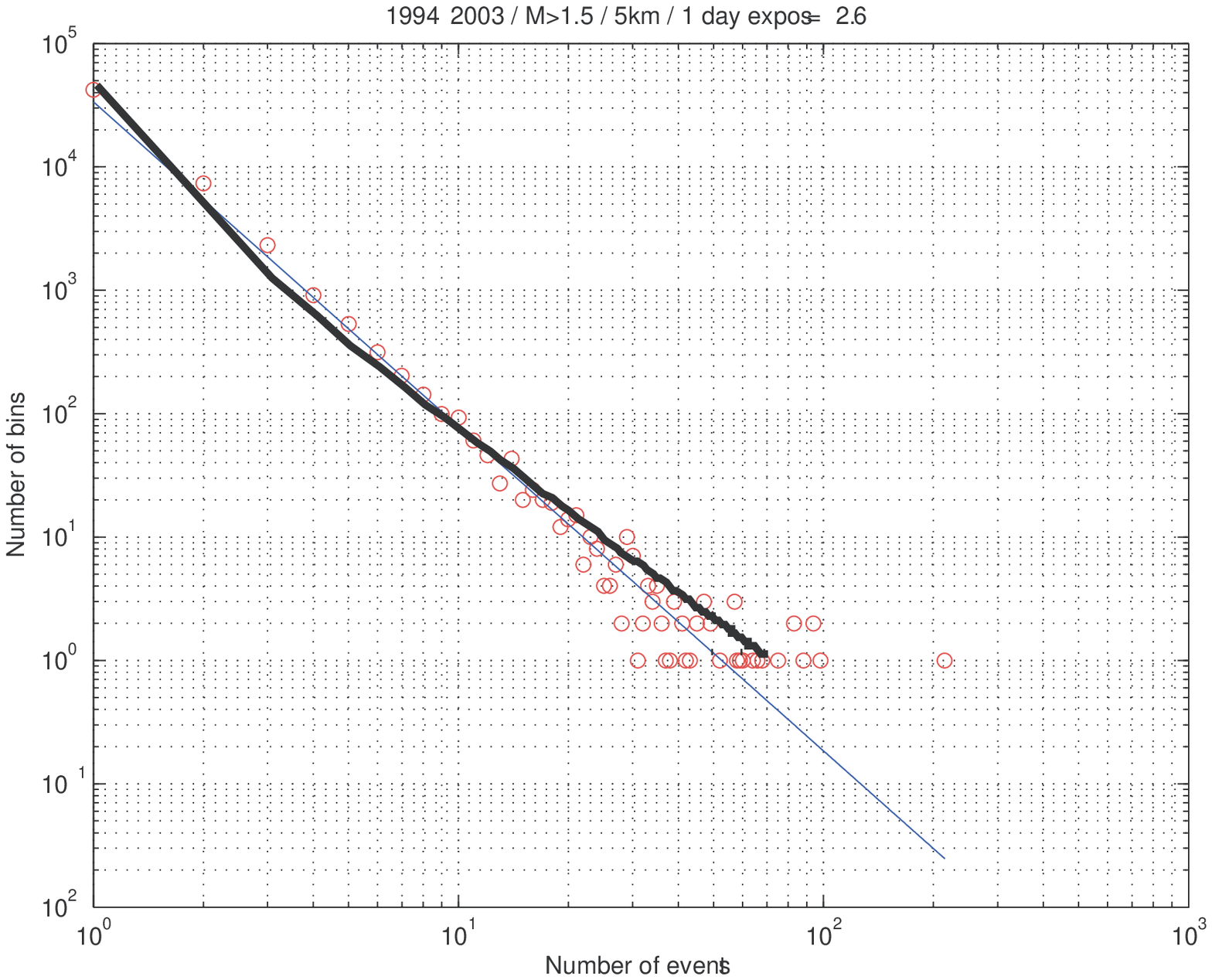}}
{\bf Fig.~10:} \small{ Empirical
probability density functions $P_{\text{data}}(r)$
of the number $r$ of earthquakes in the space-time bins of size $5 \times 5$
km$^2$ and $dt=1$ day.
The straight line is the best fit with a pure power law
$P_{\text{data}}(r) \sim 1/r^{1+\mu}$ over the range $10 \leq r \leq
100$, which
gives $\mu = 1.60$. The continuous line is the theoretical prediction using
formula (\ref{fjjalala}) with $dt=1$ day and with no
adjustable parameters, as the parameters are fixed to the values adjusted
from the fit in Fig.7.}
\end{quote}

\clearpage
\begin{psfrags}
\psfrag{P}{$P_k(r)$} \psfrag{k=1}{$k=1$}
\psfrag{k=inf}{$k=\infty$}
\psfrag{r}{$r$}
\begin{quote}
\centerline{
\resizebox{16cm}{!}{\includegraphics{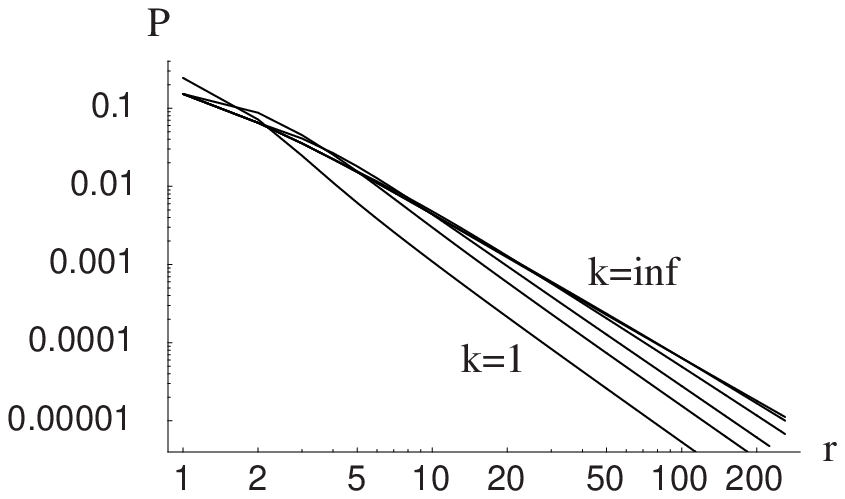}}} {\bf
Fig.~11:} \small{Distributions $P_k(r)$ given by
(\ref{53}) for $\gamma=1.25$, $n=1$, for different
numbers of generations $k=1;\, 2;\, 3;\, 5;\, 8$. The
upper curve corresponds to asymptotic distribution
$P(r)$ given (\ref{54}) corresponding to $k \to
+\infty$, while lower one corresponds to the
distribution $P_1(r)$ given by (\ref{13}) of the number
of aftershocks of first generation.}
\end{quote}
\end{psfrags}

\clearpage
\begin{psfrags}
\psfrag{p}{$p_k(r)$} \psfrag{r}{$r$}
\psfrag{k=2}{$k=2$} \psfrag{k=4}{$k=4$}
\psfrag{k=6}{$k=6$} \psfrag{k=8}{$k=8$}
\psfrag{k=10}{$k=10$}
\begin{quote}
\centerline{
\resizebox{16cm}{!}{\includegraphics{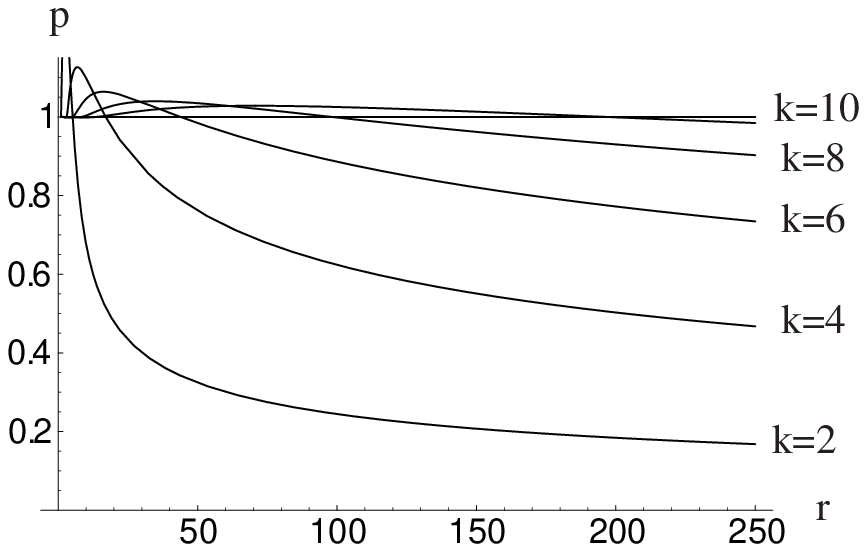}}} {\bf
Fig.~12:} \small{Ratios (\ref{55}) as a function of
event numbers $r$ for different values of the generation
number $k$, demonstrating the convergence of the
distributions $P_k(r)$ to the asymptotic distribution
$P(r)$. Bottom to top: $k=2;\, 4;\, 6;\, 8;\, 10$.}
\end{quote}
\end{psfrags}

\clearpage
\begin{psfrags}
\psfrag{k}{$k$} \psfrag{p_k}{$p_k(\mathcal{S};\bs{x})$}
\psfrag{l}{$x=\ell$}
\begin{quote}
\centerline{
\resizebox{16cm}{!}{\includegraphics{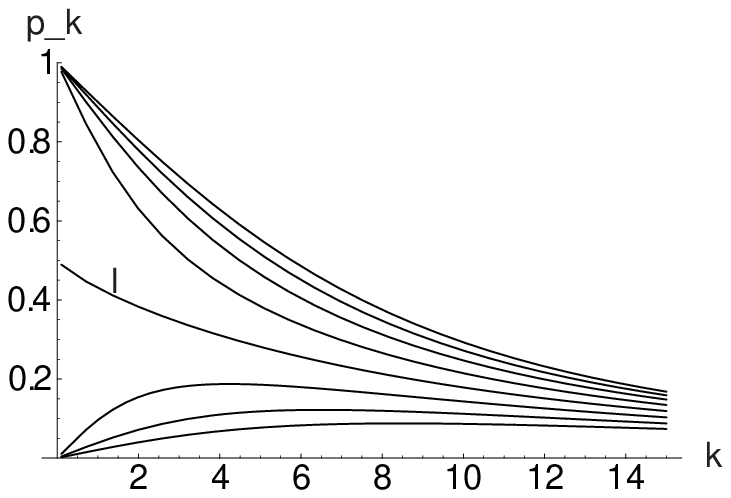}}} {\bf
Fig.~13:} \small{Plots of the probabilities $P_k(r,\mathcal{S};\bs{x})$
given by (\ref{61}) for $\gamma=1.25$, $n=0.99$,
$\ell=10 d$ and for different positions of the mother earthquake:
$x/\ell=0;~0.4; ~0.6; ~0.8; ~1; ~1.2;~1.4;~1.6$.
Recall that $\ell$ is the radius of the assumed circular domain
$\mathcal{S}$ centered on the origin. The two families of curves
separated by the central one for $x/\ell=1$ are explained in the text.}
\end{quote}
\end{psfrags}

\clearpage
\begin{psfrags}
\psfrag{P}{$P(r,\mathcal{S};\bs{x}=0)$} \psfrag{r}{$r$}
\psfrag{l=d}{$\ell=d$}
\psfrag{l=20d}{$\ell=20d$}
\begin{quote}
\centerline{
\resizebox{16cm}{!}{\includegraphics{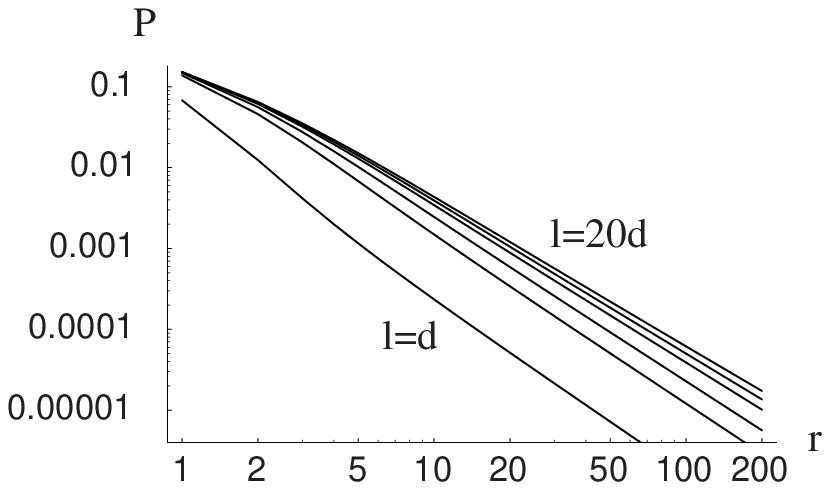}}} {\bf
Fig.~14:} \small{Plots of the distribution
$P(r,\mathcal{S};\bs{x})$ of the total number $r$ of
aftershocks falling within the disk $\mathcal{S}$ for a mother
earthquake at the origin $\bs{x}=0$ and for different values of the
circle radius
$\ell$. Bottom to top: $\ell/d=1;\,3;\,5;\,10;\,20$.}
\end{quote}
\end{psfrags}

\clearpage
\begin{psfrags}
\psfrag{P}{$P(r,\mathcal{S};\bs{x})$} \psfrag{r}{$r$}
\psfrag{}{$$}
\begin{quote}
\centerline{
\resizebox{16cm}{!}{\includegraphics{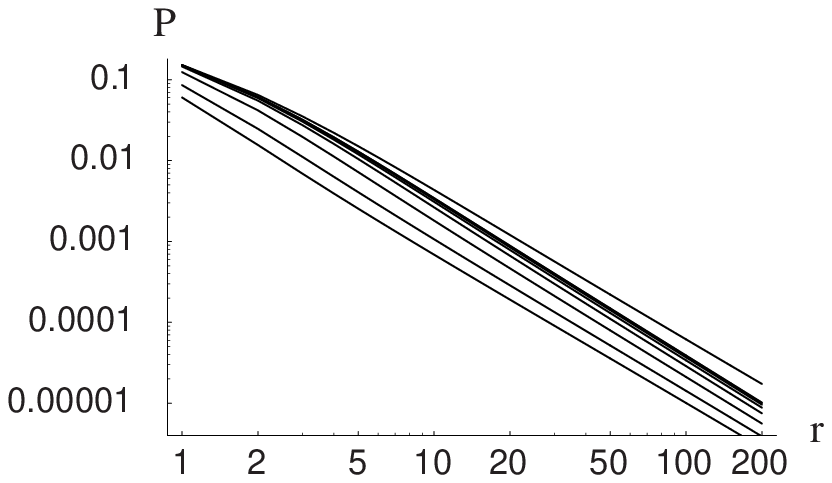}}} {\bf
Fig.~15:} \small{Plots of the distribution
$P(r,\mathcal{S};\bs{x})$ of the total number $r$ of
aftershocks falling within the disk $\mathcal{S}$ for
different positions $\bs{x}$ of the mother earthquake at fixed
disk radius $\ell/d=10$. Bottom to top: $x/\ell=1.4;\, 1.2;\, 1;\,
0.8;\, 0.6;$ $\, 0.4;\,
0.2;\, 0$. The upper curve thus corresponds to an infinite disk
$\ell= +\infty$.}
\end{quote}
\end{psfrags}

\clearpage
\begin{psfrags}
\psfrag{P}{$P(r)$} \psfrag{r}{$r$}
\psfrag{l=10d}{$\ell=10d$} \psfrag{l=5d}{$\ell=5d$}
\psfrag{l=7.5d}{$\ell=7.5d$}
\psfrag{l=2.5d}{$\ell=2.5d$}
\psfrag{p=0.6}{$p=0.6$} \psfrag{p=0.55}{$p=0.55$}
\psfrag{p=0.4}{$p=0.4$} \psfrag{p=0.18}{$p=0.18$}
\begin{quote}
\centerline{
\resizebox{16cm}{!}{\includegraphics{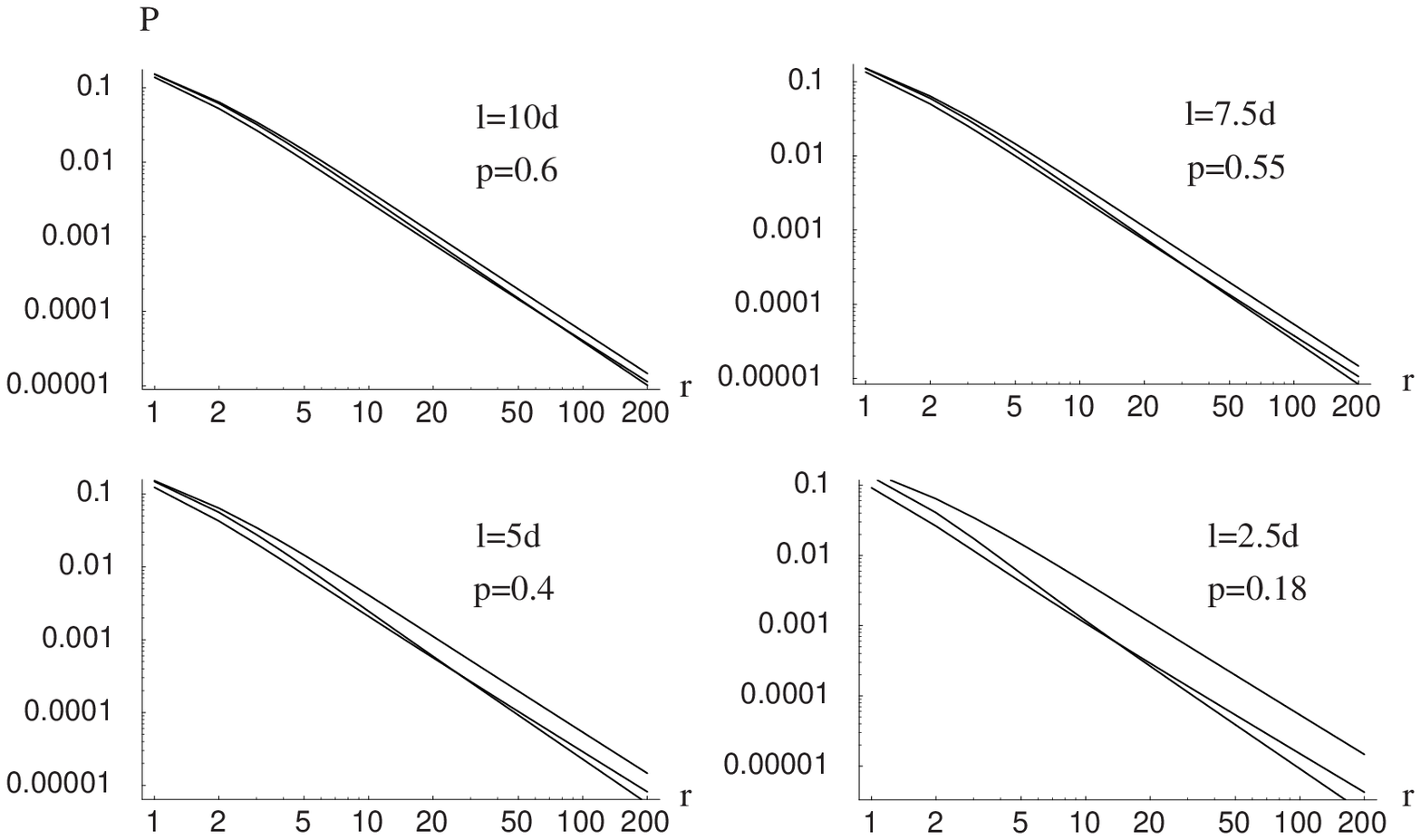}}} {\bf
Fig.~16:} \small{Comparison of the asymptotic distribution
$P(r,\mathcal{S},\bs{x}=0)$, obtained by calculating the
integral (\ref{60a}) for a large enough generation number $k$ ($k=25$
is found to be sufficient), with the factorization approximation
$P(r,p)$ given by (\ref{fact}), where $p = p_\mathcal{S}$ is defined
as the fraction of the aftershocks which fall into the domain $\mathcal{S}$.
The upper curve in each panel is
the distribution (\ref{54}) for an infinite domain $\ell/d = +\infty$,
as a reference. The different panels correspond to
$\ell/d = 10; ~7.5; ~5; ~2.5$, with
$\gamma=1.25$, $n=0.99$, $\bs{x}=0$.}
\end{quote}
\end{psfrags}

\end{document}